\documentclass[10pt,aps,prd,amsmath,amssymb,superscriptaddress,twocolumn,nofootinbib,showpacs,preprintnumbers,amsfonts,notitlepage,floatfix]{revtex4-2}

\usepackage{graphicx}
\usepackage{multirow}
\usepackage{amsmath,amssymb,amsfonts}
\usepackage{amsthm}
\usepackage{mathrsfs}
\usepackage[title]{appendix}

\usepackage{booktabs}
\usepackage{adjustbox}
\usepackage{orcidlink}

\usepackage{array}

\usepackage{siunitx}
\renewcommand{\arraystretch}{1.5}

\usepackage{cleveref}
\Crefname{section}{Sec.}{Secs.}
\Crefname{subsection}{Sec.}{Secs.}
\Crefname{figure}{Fig.}{Figs.}
\Crefname{table}{Tab.}{Tabs.}
\begin{document}

\title{Secondary neutron beam production at Jefferson Lab}

\newcommand{\jlab}{Thomas  Jefferson  National  Accelerator  Facility,  Newport  News,  VA  23606,  USA}
\newcommand{\catania}{INFN Sezione di Catania, I-95123 Catania, Italy}
\newcommand{\genova}{INFN sezione di Genova, I-16146 Genova, Italy}
\newcommand{\unige}{Università di Genova, I-16146 Genova, Italy}
\newcommand{\messina}{Dipartimento di Scienze Matematiche e Informatiche, Scienze Fisiche e Scienze della Terra,
Universit\`a degli Studi di Messina, I-98166 Messina, Italy}

\author{A.~\surname{Fulci}\orcidlink{0000-0002-3801-725X}}
\email{antonino.fulci@unime.it}
\affiliation{\messina} 
\affiliation{\catania} 

\author{P.~\surname{Ambrozewicz}\orcidlink{0000-0001-9128-9441}}
\affiliation{\jlab}
\author{M.~\surname{Battaglieri}\orcidlink{0000-0001-5002-8771}}
\affiliation{\genova}

\author{M.~\surname{Bond\`i}\orcidlink{0000-0001-8297-9184}}
\affiliation{\catania}
\author{R.~\surname{De Vita}\orcidlink{0000-0003-2770-6751}}
\affiliation{\jlab}
\affiliation{\genova}
\author{M.~\surname{Filippini}\orcidlink{0009-0000-6347-1123}}
\affiliation{\messina}
\affiliation{\catania}
\author{G.~\surname{Foti}\orcidlink{0009-0000-9791-3823}}
\affiliation{\messina}
\affiliation{\catania}
\author{S.~\surname{Grazzi}\orcidlink{0000-0001-5606-4976}}
\affiliation{\genova}
\author{G.~\surname{Mandaglio}\orcidlink{0000-0003-4486-4807}}
\affiliation{\messina}
\affiliation{\catania}
\author{T.~\surname{Nagorna}\orcidlink{0009-0004-9143-2008}}
\affiliation{\genova}
\author{A.~\surname{Pilloni}\orcidlink{0000-0003-4257-0928}}
\affiliation{\messina}
\affiliation{\catania}
\author{M.~\surname{Spreafico}\orcidlink{0000-0001-5599-6931}}
\affiliation{\genova}

\author{S.~\surname{Vallarino}\orcidlink{0000-0001-6679-0728}}
\affiliation{\genova}

\begin{abstract}
We investigate secondary neutron production by an 11~GeV electron beam in the Hall~A beam dump at Jefferson Lab and their transport to a downstream experimental area. This secondary beam could operate opportunistically with the primary electron-scattering programme. The neutron energy spectrum extends from thermal energies to several GeVs, offering a distinctive opportunity for irradiation at  energy grater than 1 GeV. We considered three configurations: the current setup, the planned BDX facility, and an optimized BDX layout with a dedicated neutron extraction pipe. We also studied how radiators made of different materials and placed upstream of the beam dump increase neutron production.

\end{abstract}
\preprint{JLAB-PHY-26-5020}
\maketitle

\section{Introduction}
Intense neutron beams are essential tools across a wide range of scientific and technological disciplines, enabling studies that span fundamental physics, materials science, nuclear engineering, chemistry, biology, and applied research. Because neutrons are electrically neutral, they penetrate deeply into matter and interact primarily through the strong nuclear force, making them uniquely sensitive probes of atomic nuclei, magnetic structures, and bulk material properties. Access to high-flux neutron beams over a broad energy range, from thermal and cold neutrons to fast and relativistic neutrons, is therefore critical for both basic research and applied applications.
High-intensity neutron sources commonly rely on spallation or fission. At Oak Ridge National Laboratory, the Spallation Neutron Source (SNS) produces moderated neutrons for materials research~\cite{MASON2006955}, while the High Flux Isotope Reactor (HFIR) is a steady-state fission source supporting neutron scattering, isotope production, and materials irradiation~\cite{Howard:2025HFIR}. HFIR's in-vessel irradiation fields cover the 100~keV--1~MeV band and extend into the MeV range for neutron-damage studies~\cite{Skutnik:2023HFIR}. 
The Back-n beamline at the China Spallation Neutron Source (CSNS) likewise spans sub-MeV and higher energies, primarily for nuclear-data measurements~\cite{Chen:2019Backn}.
At higher energies, ChipIr at ISIS and the TRIUMF neutron irradiation facilities provide broad spectra extending to several hundred MeV for accelerated tests of neutron-induced single-event effects in electronics~\cite{Cazzaniga:2024ChipIr,BelangerChampagne:2024PIFNIF}. 
Together, these facilities illustrate applications across the energy intervals considered and provide the context for the comparison in \Cref{subsec:beam_comparison}.

Complementary to dedicated neutron facilities, high-current, multi-GeV electron beams interacting with thick targets can also produce intense neutron fields through electromagnetic and hadronic cascades. These environments naturally generate neutron spectra extending from thermal energies to several GeV, closely resembling conditions relevant to radiation shielding, accelerator-driven systems, space radiation environments, and detector backgrounds.
At Jefferson Lab, the electron beam from the Continuous Electron Beam Accelerator Facility (CEBAF) is absorbed in the Hall~A beam dump (BD), downstream of the targets used for hadronic physics experiments. This configuration provides a unique environment for studying neutron production and the deep penetration of neutrons through shielding materials such as concrete and soil.
In this paper, we report a study performed using a state-of-the-art simulation framework to explore the characteristics of the secondary neutron field emerging from Hall A's BD. We investigate both the standard Hall~A configuration and an optimized geometry designed to enhance the neutron flux. 

\Cref{jlab} briefly describes the CEBAF electron beam characteristics. \Cref{simulations} describes the simulation framework used to carry out the study. Neutron production and propagation in the existing and an optimized configuration are presented in \Cref{prod,prop}, respectively.

\section{Hall A at Jefferson Lab}\label{jlab}

The CEBAF at Jefferson Lab is a superconducting radio-frequency recirculating electron accelerator supporting a broad program in hadron and nuclear physics. It delivers continuous-wave polarized electron beams with energies up to 12~GeV, currents up to $\mathcal{O}(100)~\mu$A, and polarization exceeding 80\%, simultaneously to four experimental halls~\cite{Leemann:2001dg,Arrington:2021alx}. Hall~B employs the large-acceptance CLAS12 detector to study nucleon structure and hadron spectroscopy~\cite{Burkert:2020akg}, while Hall~C uses magnetic spectrometers for precision electron-scattering measurements~\cite{Arrington:2021alx}. Hall~D combines a linearly polarized photon beam with the GlueX detector to explore meson spectroscopy and search for exotic hadrons~\cite{GlueX:2020idb}.

Hall~A uses high-resolution magnetic spectrometers for precision electron-scattering studies of nucleon and nuclear structure and parity-violating observables~\cite{Alcorn:2004sb}. After passing through the experimental targets its electron beam is absorbed by a high-power BD capable of dissipating nearly 1~MW. The BD consists of approximately 80 aluminum disks (40~cm diameter) with thickness increasing from 1~cm to 2~cm over a total length of about 2~m, followed by a 50~cm diameter, 1~m long aluminum cylinder. The assembly is water-cooled to maintain stable operation under sustained high-power conditions.

The BD is surrounded by several meters of concrete shielding in both longitudinal and transverse directions, with additional soil overburden above the installation. During operation, the interaction of the multi-GeV, high-current beam with the BD generates intense electromagnetic and hadronic cascades, producing a broad-spectrum neutron field capable of penetrating deeply into the shielding. This configuration provides a well-defined and reproducible environment for studies of radiation transport, shielding performance, and secondary particle production within the routine CEBAF physics program.

The planned facility for the Beam Dump eXperiment (BDX) will be located downstream of the Hall~A BD. Designed primarily for light-dark-matter searches, this facility could also host experiments using secondary particles produced in the BD~\cite{Achenbach:2025ynn}. Here, we study the Hall~A BD as a neutron source and assess the neutron flux that could reach the planned BDX experimental area during routine electron-beam operation.

\section{The Simulation Framework}\label{simulations}
\begin{figure}[!htbp]
\centering
\includegraphics[width=\linewidth]{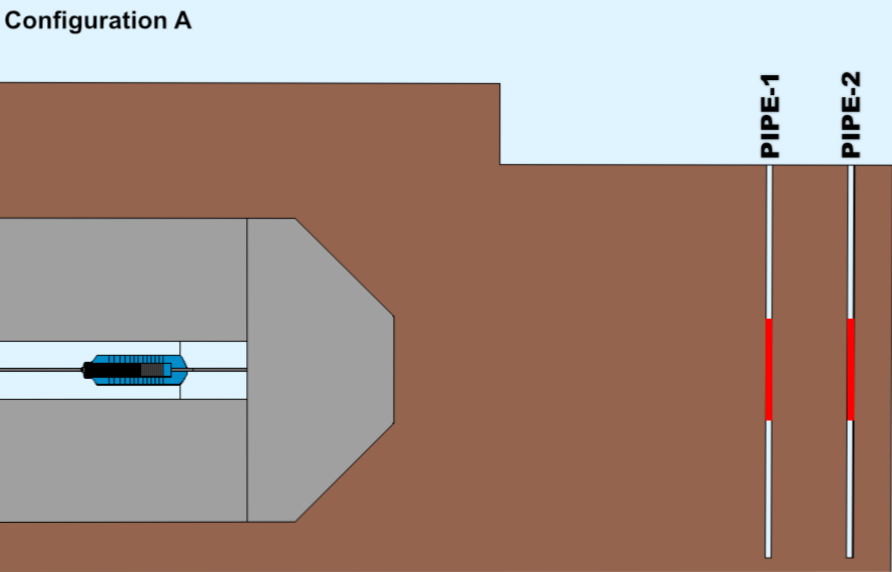}\\
\includegraphics[width=\linewidth]{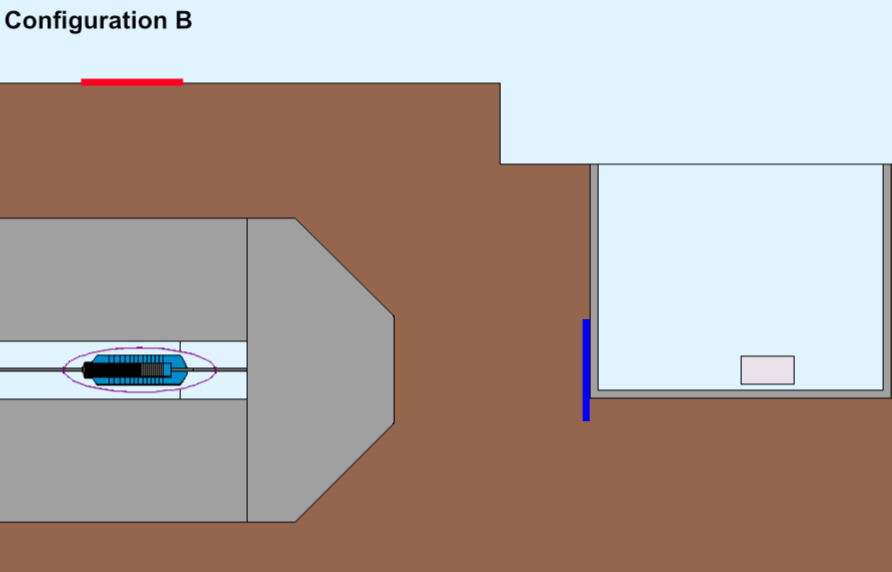}\\
\includegraphics[width=\linewidth]{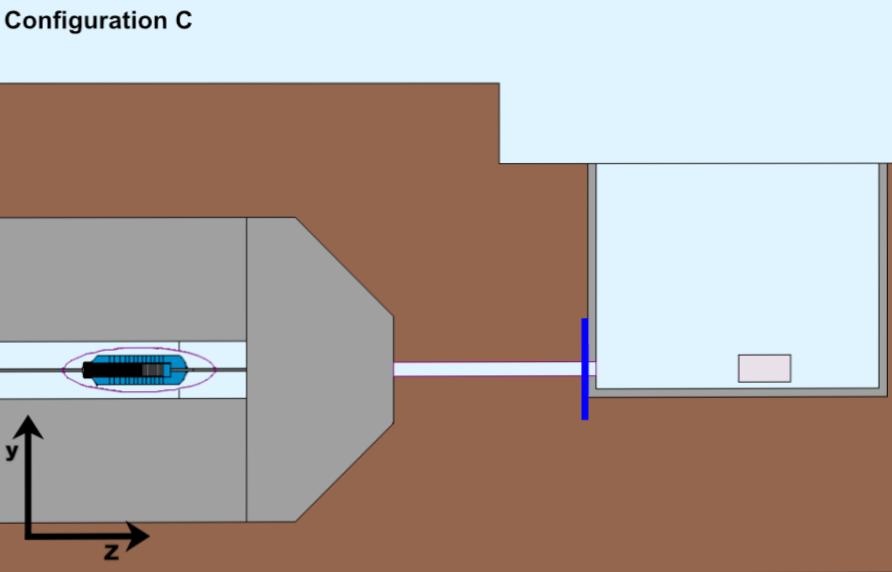}
\caption{Cross-sectional views ($yz$ plane) of configurations A, B, and C, shown from top to bottom. The drawings show the Hall~A BD (blue), the concrete vault (gray), the surrounding dirt (brown), and air (light blue). Configuration A shows the current setup with the two BDX-MINI experiment wells. Configuration B shows the planned BDX Facility with the BDX detector (light pink). Configuration C shows the horizontal air-filled pipe connecting the BD concrete vault to the BDX Facility to enhance the neutron flux at the facility entrance. The colored surfaces indicate the neutron sampling regions used in the simulations.}
\label{fig:geom}
\end{figure}

\begin{figure}[t]
\centering
\includegraphics[width=0.8\linewidth]{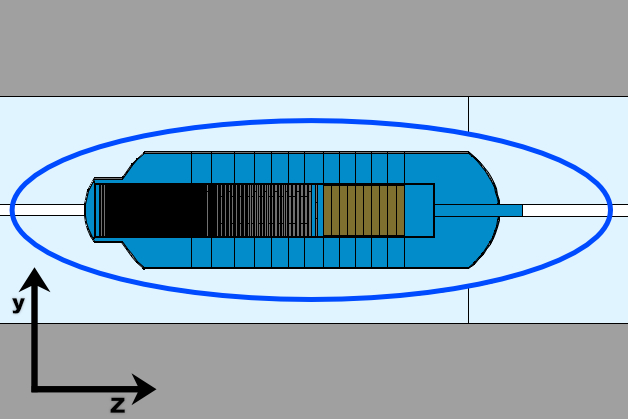}
\caption{Cross-sectional view ($yz$ plane) of the Hall~A BD as implemented in FLUKA. The drawing shows the aluminum disks (amber), water (blue), air (light blue), and concrete (gray). The blue outline indicates the ellipsoidal $4\pi$ surface enclosing the BD used to sample neutrons at production.}
\label{fig:bd_geom}
\end{figure}

We modeled the interaction of JLab's $11$~GeV electron beam with the Hall A BD, and the subsequent production and propagation of neutrons through concrete and soil using the FLUKA version 4-5.2~\cite{fluka-cern, Ahdida:2022gjl, Battistoni:2015epi}.
We consider the beam as an ideal pencil beam: monoenergetic at $11$~GeV, with zero emittance, impinging at the center of the upstream face of the BD.
The BD, the beam transport line, and the surrounding concrete vault are shown (to scale) in \Cref{fig:geom}, while \Cref{fig:bd_geom} shows a close-up cross-sectional view of the simulated BD geometry and the neutron sampling surface.
We validated the geometry against engineering drawings of the Hall A complex~\cite{JLabDrawingS203, JLabDrawingS214, JLabDrawing27020E0001, JLabDrawing27020E0139}.
In this work we consider three configurations:
\begin{itemize}
    \item \textbf{Configuration A}: the current \emph{as-built} setup, including the two wells dug for the BDX-MINI experiment~\cite{Battaglieri:2022dcy};
    \item \textbf{Configuration B}: setup including the proposed BDX Facility as described in~\cite{Achenbach:2025ynn}, expected to be deployed in the near future;
    \item \textbf{Configuration C}: neutron-extraction configuration featuring a 50.8~cm (20~in) diameter horizontal air-filled pipe connecting the BD concrete vault to the BDX Facility. We simulate this configuration both with a \mbox{high-Z} pre-BD target, to enhance the low-energy neutron yield, and without it, to preserve the high-energy component.
\end{itemize}

The baseline neutron-production geometry is common to all three configurations. The \mbox{high-Z} pre-target is studied as an independent option in configurations~B and~C, whereas the extraction pipe in configuration~C modifies downstream transport.

The simulation setup is based on the FLUKA \texttt{PRECISIO} defaults, with explicit settings for nuclear interactions, low-energy neutron transport, particle-production thresholds, and variance reduction. Photonuclear interactions are enabled over all implemented energy regimes through \texttt{PHOTONUC}, together with electronuclear interactions through its \texttt{ELECTNUC} option. Neutron transport below 20~MeV uses the pointwise treatment with the JEFF-3.3 evaluated nuclear-data library~\cite{Plompen:2020JEFF33}, selected through \texttt{LOW-PWXS}. Coalescence is enabled through \texttt{PHYSICS}. The evaporation settings use the model with heavy-fragment evaporation. These settings supplement the general FLUKA physics configuration described in Refs.~\cite{Ahdida:2022gjl,Battistoni:2015epi}.

Two families of simulation runs were prepared: an ordinary family covering a broad neutron-energy range and a second family with higher transport cutoffs, intended to improve computational efficiency when sampling the high-energy neutron component.
In the ordinary propagation runs, the electromagnetic thresholds are set to 10~MeV to retain lower-energy electromagnetic cascades and their contribution to secondary neutron production, while the default neutron cutoff of $10^{-5}$~eV is retained.
In the second family, the electromagnetic thresholds are raised to 0.1~GeV and a neutron kinetic-energy cutoff of approximately 400~MeV is applied, suppressing lower-energy transport in both components.

Photonuclear interactions in the aluminum alloy and water of the BD are preferentially sampled using \texttt{LAM-BIAS}, with an interaction-length scaling factor of 0.005. FLUKA adjusts particle weights to account for the modified interaction sampling. The increased number of sampled interactions therefore improves access to rare neutron-production processes without representing a corresponding increase in the physical source intensity. The specified material scope is retained explicitly, rather than extending this biasing prescription to every component of the geometry.

Leading Particle Biasing (LPB) is additionally configured through \texttt{EMF-BIAS} with the \texttt{LPBEMF} option. It reduces the computational cost of electromagnetic cascades by selectively retaining particles from electromagnetic interactions and compensating their selection probabilities through statistical weights.

Deep-penetration calculations employ user-defined importance biasing. The routine returns an importance ratio based on material, step length, position, and direction relative to a selected target point. FLUKA uses this ratio for particle splitting and Russian roulette, with the corresponding weight adjustments.

All neutron observables are evaluated using the statistical weights carried by the scored crossings.
Standard variance-reduction principles are discussed in Ref.~\cite{10.3389/fphy.2021.718873}.

The number of simulated primary electrons varies by configuration and family run. For the neutron production studies at the BD approximately $10^7$ electrons on target (EOT) were simulated, while the high-energy no-radiator sample comprised $10^8$~EOT. Propagation samples span approximately $10^8$--$10^9$~EOT.

A custom Fortran--C++ interface records the position, direction cosines, kinetic energy, and statistical weight of each neutron crossing a scoring surface in a \texttt{ROOT} file~\cite{FlukaROOTOutput}.
This unbinned output allows spatial, angular, and energy selections and histogram binning to be changed offline without rerunning the simulation.
A dedicated analysis framework developed for this work handles crossing selection, energy-spectrum reconstruction, curve smoothing, and rate calculations~\cite{BDXFlukaAnalysis}.

\section{Neutron Production at the BD}\label{prod}

When a high-energy electron beam impinges on a thick target, an intense electromagnetic shower develops, producing a wide spectrum of secondary particles. Electrons lose energy predominantly through bremsstrahlung, generating high-energy photons that, in turn, interact with nuclei via photonuclear reactions leading to neutron emission. 
This mechanism represents the primary source of neutrons in thick high-Z targets irradiated by electron beams~\cite{ParticleDataGroup:2022pth}.

The electromagnetic character of the initiating shower should be distinguished from the subsequent nuclear dynamics. Neutrons can be emitted following photon absorption below the pion-production threshold ($\sim$140 MeV), while above it additional channels involving pions and baryon resonances contribute to the coupled electromagnetic and hadronic shower. This behavior is implemented in FLUKA and similar Monte Carlo frameworks. ~\cite{Battistoni:2015epi, Fasso1994, Ranft:1986dc}.

The photonuclear response involves overlapping energy regimes rather than mutually exclusive reaction classes:
\begin{itemize} 
    \item \textbf{Giant Dipole Resonance}, dominant for photon energies of the order 10-30~MeV, where collective nuclear excitations lead to the emission of one or two neutrons~\cite{BertulaniDanielewicz2021};
    \item \textbf{Quasi-deuteron absorption}, effective between $30$ MeV and a few hundred MeV, where photons are more likely to interact with a neutron-proton pair rather than with all nucleons collectively~\cite{Levinger:1951vp};
    \item \textbf{Baryon-resonance and meson-production channels}, occurring between approximately $140$ MeV and $1.2$ GeV; the $\Delta(1232)$ is prominent at photon energies of a few hundred MeV; higher resonances and additional meson-production channels contribute as the photon energy increases towards the GeV range.~\cite{Koch:1983tf,Ericson:1988gk};
    \item \textbf{High-energy photonuclear interactions}: the hadronic component of the photon can be described using vector-meson-dominance models, with the resulting interaction also depending on the nuclear target. ~\cite{Bauer:1977iq};
\end{itemize}

Following the initial interaction, intranuclear rescattering, pre-equilibrium emission and nuclear de-excitation can contribute to the outgoing neutron population. Evaporation and, where relevant, fission are subsequent reaction stages rather than separate photon-energy bands.~\cite{Battistoni:2015epi, Iljinov1994}.

In the Hall~A BD, the 11~GeV CEBAF electron beam initiates electromagnetic and hadronic showers within an aluminum-water composite target. These showers propagate throughout the BD's volume, generating a neutron spectrum that spans from thermal energies to several GeV. While high-energy neutrons are strongly peaked in the forward direction due to the kinematics of photonuclear reactions induced by high-energy Bremsstrahlung photons, the low-energy component exhibits a different angular profile. Although nuclear evaporation and scattering processes are inherently near-isotropic, the low-energy flux is significantly enhanced in the backward direction. This is a consequence of the `albedo' effect from the BD face and the moderation of neutrons that undergo multiple large-angle scattering events within the surrounding shielding and the materials of the BD structure, especially when high-Z components are placed before the BD~\cite{schaeffer_2025_xr8p6-q9r07}.

To estimate the neutron yield produced by the BD, we sampled the neutron flux on a closed surface fully enclosing the BD volume, providing complete $4\pi$ angular coverage, as illustrated in \Cref{fig:bd_geom}.
For configurations A, B and C (without the pre-BD target), the resulting integrated production yield amounts to approximately $Y_n \simeq 0.11 \,\text{n/EOT}$. This baseline production will be used as a reference for evaluating the impact of high-$Z$ materials introduced in Configuration~C.

In fact, the production of secondary neutrons can be significantly enhanced by inserting a high-$Z$ material immediately upstream of the aluminum-water core of the BD. 
Such inserts act as efficient photon converters, generating an increased flux of Bremsstrahlung photons which, in turn, undergo photonuclear interactions, predominantly enhancing neutron production below $\sim$100~MeV.\\
To quantify this effect, we investigated different materials by placing a circular target (radius $R=10$~cm) in front of the BD. We start investigating a thickness $\Delta z=5$~cm.

We sampled the neutron flux again on a $4\pi$ surface surrounding the BD for each configuration, allowing a direct comparison with the unmodified baseline. 
To quantify the relative increase in neutron production, an enhancement factor (EF) is defined as the ratio of the yield for each radiator configuration to the no-radiator baseline, $\mathrm{EF} = Y / Y_{\mathrm{None}}$. The corresponding yields and enhancement factors are reported in \Cref{tab:neutron_yields}.

\begin{table}[t]
\centering
\small
\renewcommand{\arraystretch}{1.15}
\setlength{\tabcolsep}{4pt}
\begin{tabular}{@{}lcc@{}}
\toprule
\textbf{Material} & \textbf{Yield (n/EOT)} & \textbf{EF} \\
\midrule
None & $0.11$ & $1.0$ \\
Pb & $2.2$ & $20.1$ \\
Hg & $2.5$ & $23.5$ \\
\textbf{W} & $\boldsymbol{3.3}$ & $\boldsymbol{30.5}$ \\
U & $5.6$ & $51.8$ \\
\bottomrule
\end{tabular}
\caption{Total neutron yield per EOT for different pre-BD radiator materials of 5~cm thickness. Statistical uncertainties are below 0.5\%.}
\label{tab:neutron_yields}
\end{table}
\begin{figure}[b]
\centering
\includegraphics[width=\columnwidth]{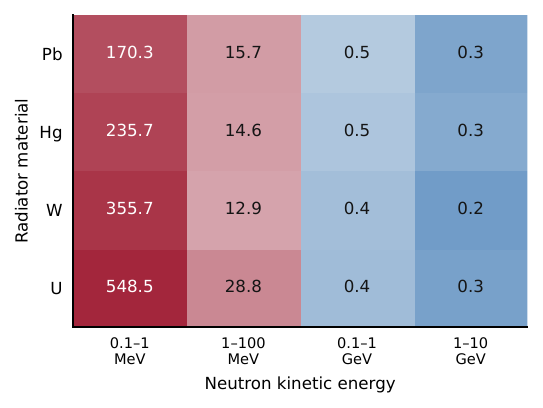}
\caption{Neutron production enhancement factors for 5~cm-thick, 10~cm-radius Pb, Hg, W and U radiators upstream of the BD. Each cell gives the ratio of the yield integrated over the indicated kinetic-energy interval to the no-radiator yield. Colors follow a logarithmic scale from EF~$=0.1$ to $1000$, with white at EF~$=1$; red denotes enhancement and blue suppression.}
\label{fig:material_enhancement}
\end{figure}

\begin{figure}[t]
\centering
\includegraphics[width=\columnwidth]{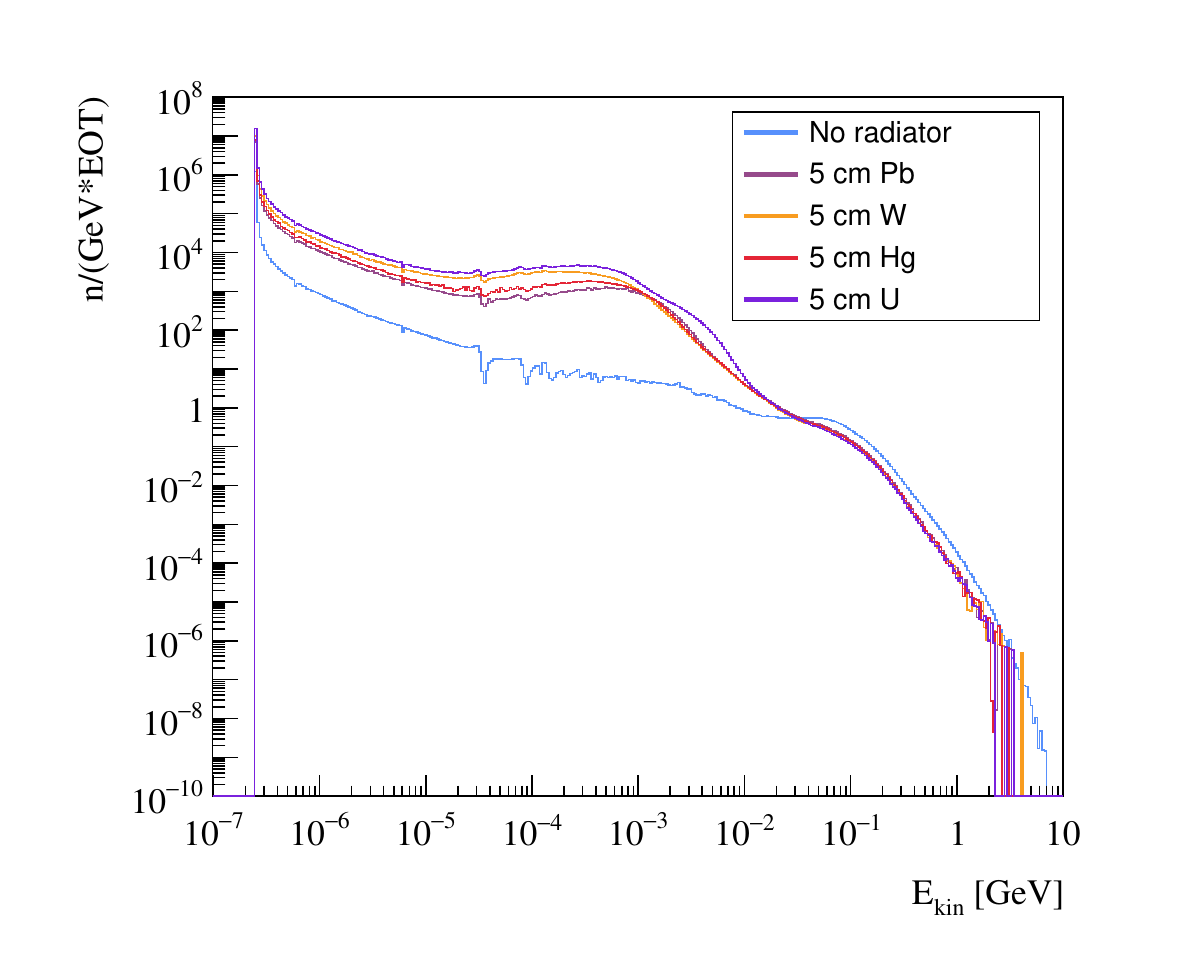}
\caption{Neutron energy spectra at production, sampled on the ellipsoidal $4\pi$ surface shown in \Cref{fig:bd_geom}, comparing the no-radiator baseline (blue) with 5~cm-thick high-$Z$ radiators.}
\label{fig:opt_spectra}
\end{figure}

\begin{table*}[t]
\centering
\small
\renewcommand{\arraystretch}{1.15}
\setlength{\tabcolsep}{4pt}
\begin{tabular}{@{}l*{4}{cc}@{}}
\toprule
\textbf{Material} & \multicolumn{2}{c}{\textbf{100~keV--1~MeV}} & \multicolumn{2}{c}{\textbf{1--100~MeV}} & \multicolumn{2}{c}{\textbf{100~MeV--1~GeV}} & \multicolumn{2}{c}{\textbf{1--10~GeV}} \\
\cmidrule(lr){2-3}\cmidrule(lr){4-5}\cmidrule(lr){6-7}\cmidrule(lr){8-9}
 & Yield (n/EOT) & EF & Yield (n/EOT) & EF & Yield (n/EOT) & EF & Yield (n/EOT) & EF \\
\midrule
None & $5.7 \times 10^{-3}$ & $1.0$ & $6.3 \times 10^{-2}$ & $1.0$ & $1.8 \times 10^{-2}$ & $1.0$ & $5.6 \times 10^{-5}$ & $1.0$ \\
Pb & $9.8 \times 10^{-1}$ & $170.3$ & $9.8 \times 10^{-1}$ & $15.7$ & $8.9 \times 10^{-3}$ & $0.5$ & $1.6 \times 10^{-5}$ & $0.3$ \\
Hg & $1.4$ & $235.7$ & $9.2 \times 10^{-1}$ & $14.6$ & $8.3 \times 10^{-3}$ & $0.5$ & $1.7 \times 10^{-5}$ & $0.3$ \\
\textbf{W} & $\boldsymbol{2.0}$ & $\boldsymbol{355.7}$ & $\boldsymbol{8.1} \times \boldsymbol{10^{-1}}$ & $\boldsymbol{12.9}$ & $\boldsymbol{7.5} \times \boldsymbol{10^{-3}}$ & $\boldsymbol{0.4}$ & $\boldsymbol{1.4} \times \boldsymbol{10^{-5}}$ & $\boldsymbol{0.2}$ \\
U & $3.1$ & $548.5$ & $1.8$ & $28.8$ & $7.0 \times 10^{-3}$ & $0.4$ & $1.4 \times 10^{-5}$ & $0.3$ \\
\bottomrule
\end{tabular}
\caption{Neutron yields for different 5~cm pre-BD radiator materials for different energy bins. For each energy bin, the two sub-columns give the neutron yield (n/EOT) and the enhancement factor (EF) relative to the no-radiator configuration. Relative statistical uncertainties are below 0.5\% except for the last energy bin, where they are below 10\%. We do not report neutrons below 100 keV.
}
\label{tab:neutron_yields_2}
\end{table*}
\begin{table*}[t]
\centering
\small
\renewcommand{\arraystretch}{1.15}
\setlength{\tabcolsep}{4pt}
\begin{tabular}{@{}c*{4}{cc}@{}}
\toprule
\textbf{$t$ (cm)} & \multicolumn{2}{c}{\textbf{100~keV--1~MeV}} & \multicolumn{2}{c}{\textbf{1--100~MeV}} & \multicolumn{2}{c}{\textbf{100~MeV--1~GeV}} & \multicolumn{2}{c}{\textbf{1--10~GeV}} \\
\cmidrule(lr){2-3}\cmidrule(lr){4-5}\cmidrule(lr){6-7}\cmidrule(lr){8-9}
 & Yield (n/EOT) & EF & Yield (n/EOT) & EF & Yield (n/EOT) & EF & Yield (n/EOT) & EF \\
\midrule
None & $5.7 \times 10^{-3}$ & $1.0$ & $6.3 \times 10^{-2}$ & $1.0$ & $1.8 \times 10^{-2}$ & $1.0$ & $5.6 \times 10^{-5}$ & $1.0$ \\
0.1 & $6.9 \times 10^{-3}$ & $1.2$ & $6.7 \times 10^{-2}$ & $1.1$ & $1.8 \times 10^{-2}$ & $1.0$ & $5.6 \times 10^{-5}$ & $1.0$ \\
0.5 & $2.6 \times 10^{-2}$ & $4.5$ & $1.0 \times 10^{-1}$ & $1.6$ & $1.7 \times 10^{-2}$ & $1.0$ & $4.4 \times 10^{-5}$ & $0.8$ \\
1.0 & $1.1 \times 10^{-1}$ & $19.7$ & $1.9 \times 10^{-1}$ & $3.0$ & $1.5 \times 10^{-2}$ & $0.8$ & $3.1 \times 10^{-5}$ & $0.6$ \\
\textbf{5.0} & $\boldsymbol{2.0}$ & $\boldsymbol{355.7}$ & $\boldsymbol{8.1} \times \boldsymbol{10^{-1}}$ & $\boldsymbol{12.9}$ & $\boldsymbol{7.5} \times \boldsymbol{10^{-3}}$ & $\boldsymbol{0.4}$ & $\boldsymbol{1.4} \times \boldsymbol{10^{-5}}$ & $\boldsymbol{0.2}$ \\
\bottomrule
\end{tabular}
\caption{Neutron yields per EOT for different W radiator thicknesses, evaluated in several neutron energy intervals. For each energy bin, the two sub-columns give the neutron yield (n/EOT) and the enhancement factor (EF) relative to the no-radiator configuration. Relative statistical uncertainties are below 0.5\% in the first three energy bins and range from 0.66\% to 8.57\% in the 1--10~GeV bin. We do not report neutrons below 100 keV.
}
\label{tab:neutron_yields_thickness}
\end{table*}

\begin{figure*}[!htbp]
\centering
\includegraphics[width=0.44\linewidth]{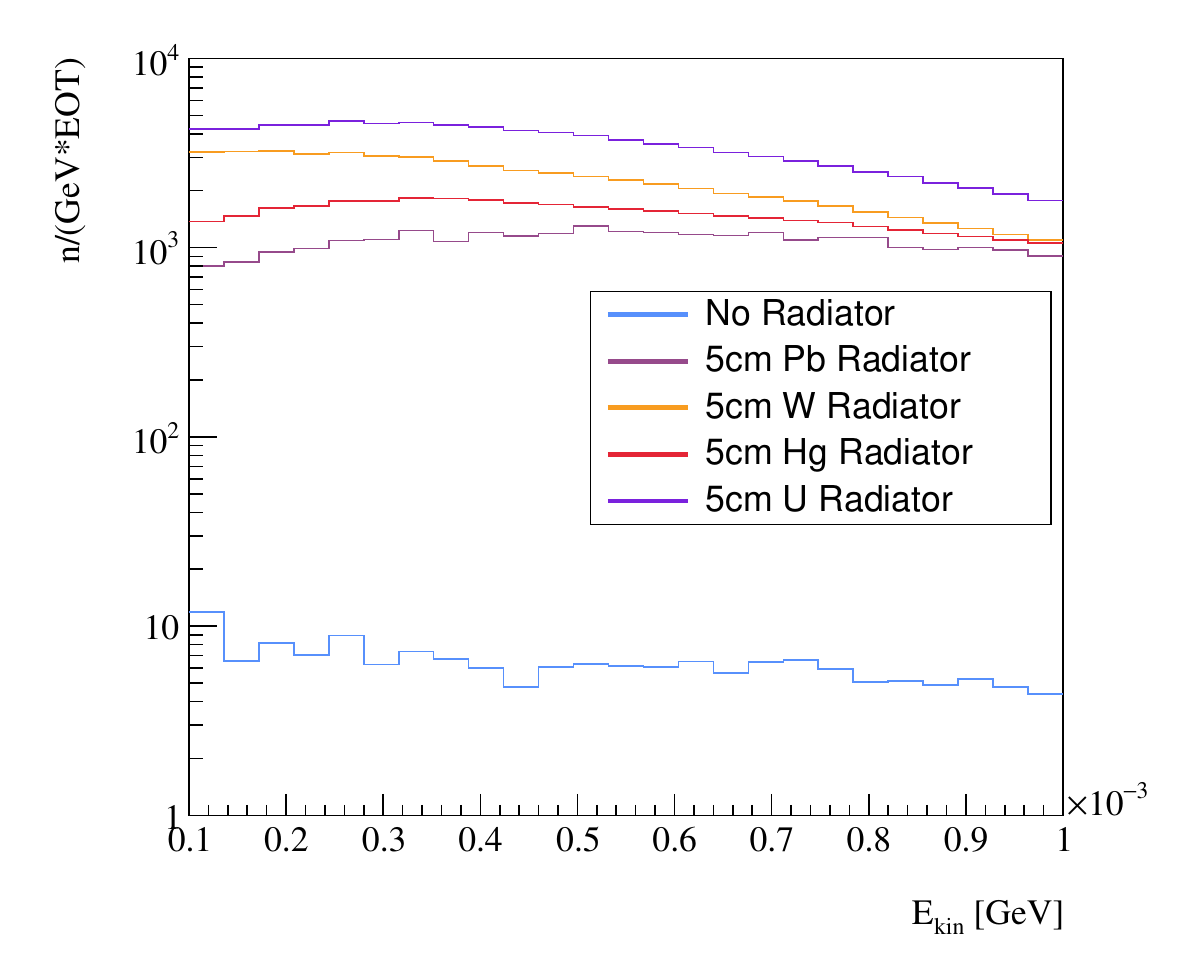}
\includegraphics[width=0.44\linewidth]{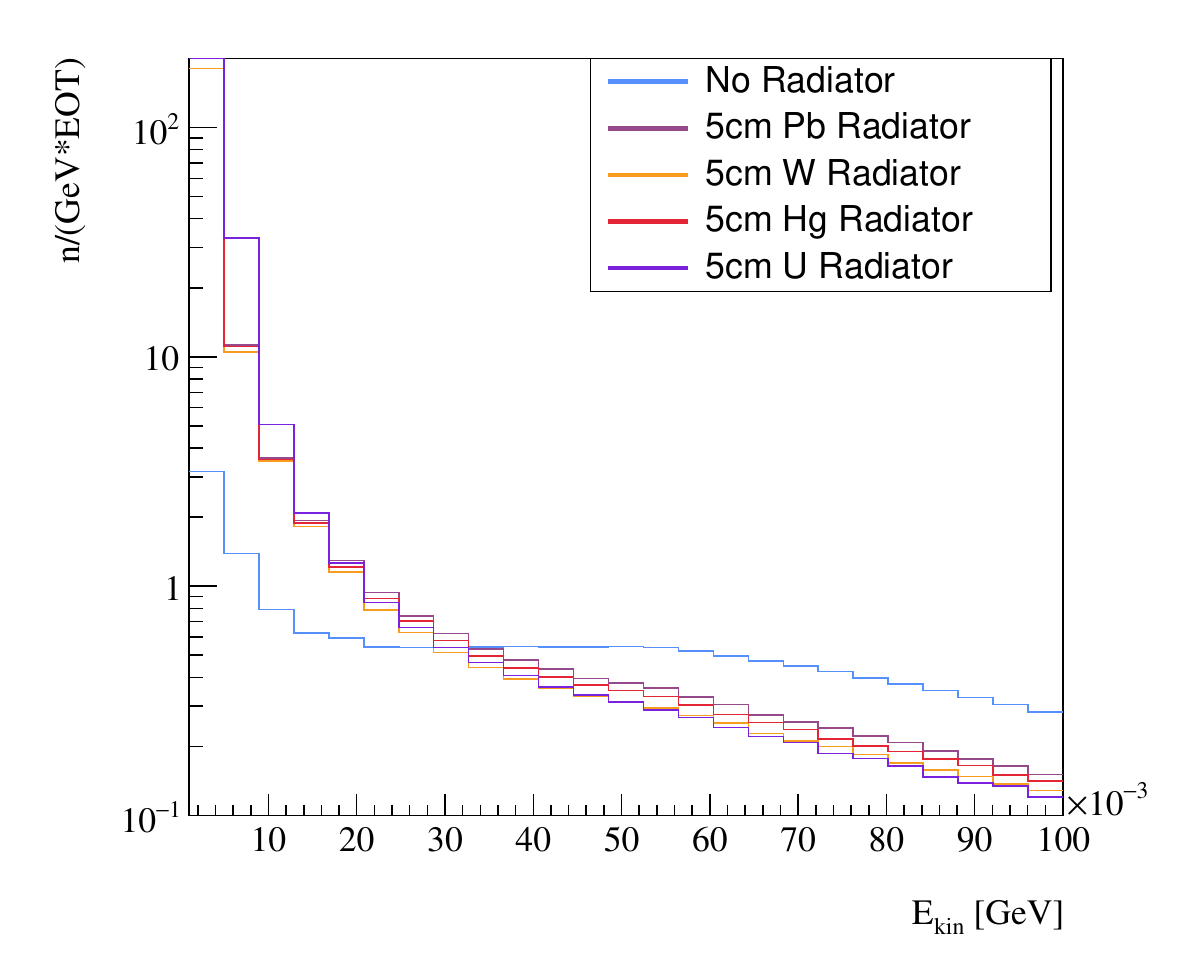}
\includegraphics[width=0.44\linewidth]{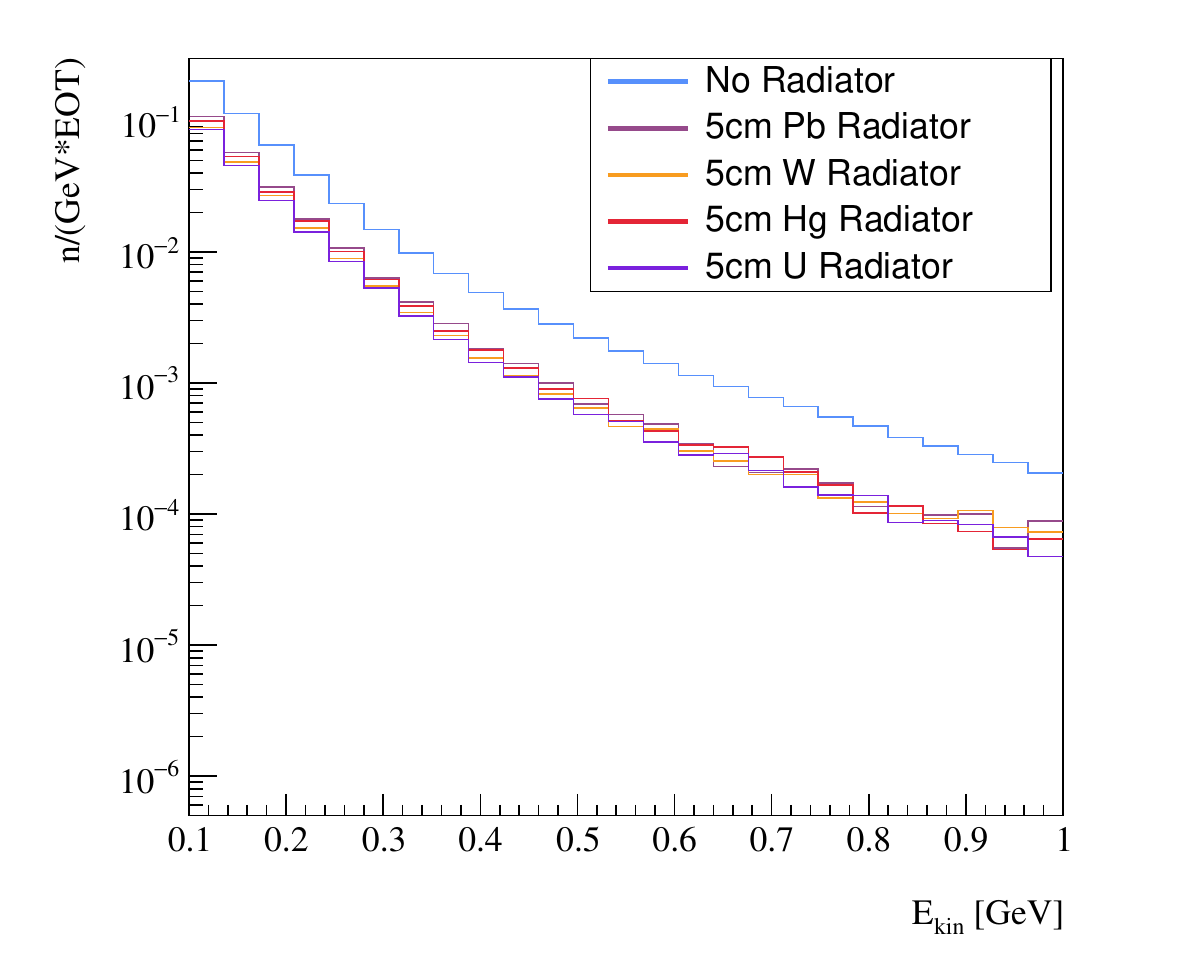}
\includegraphics[width=0.44\linewidth]{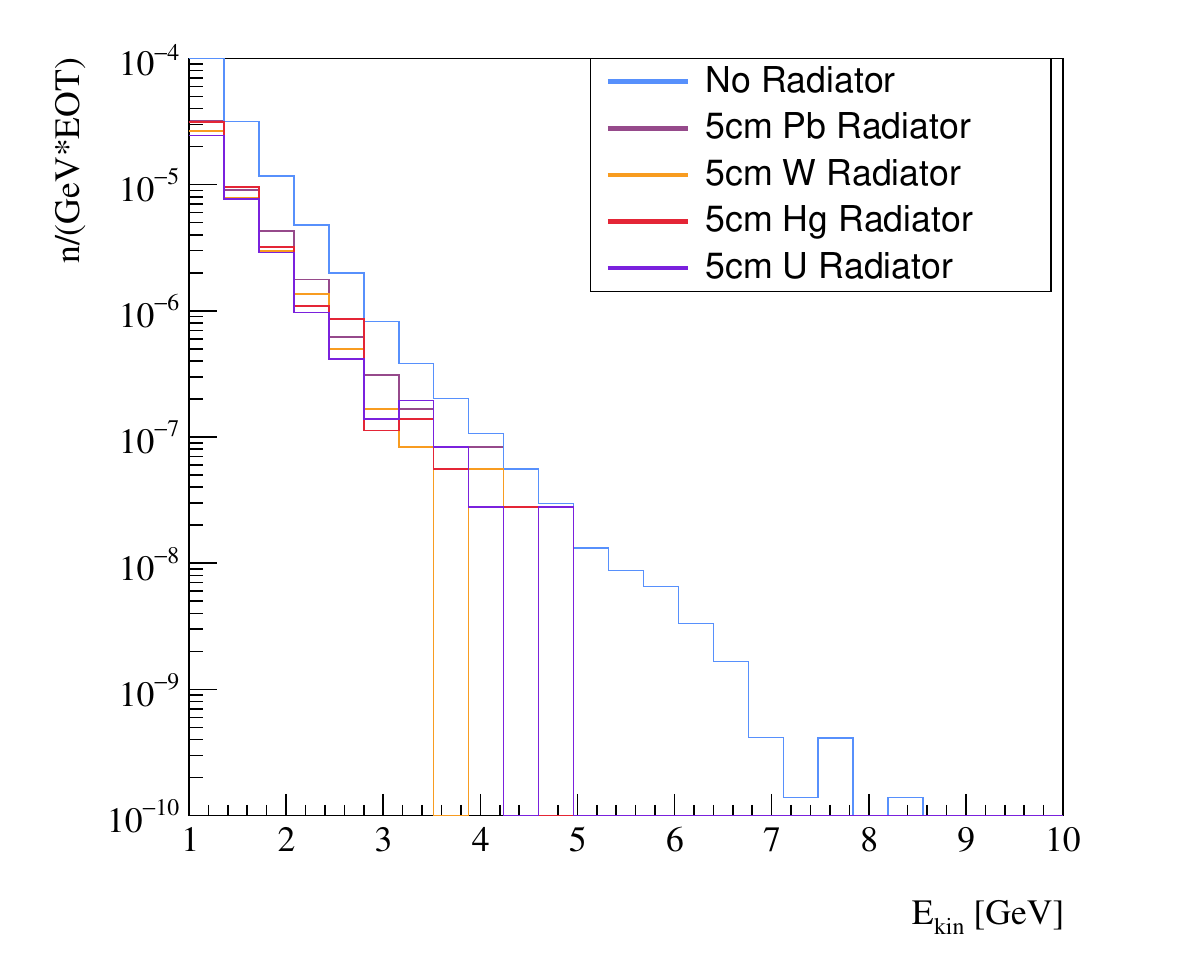}
\caption{Neutron energy spectra at production, sampled on the ellipsoidal $4\pi$ surface shown in \Cref{fig:bd_geom}, in different energy ranges. \textbf{Top-left}: 100 keV--1 MeV, \textbf{top-right}: 1 MeV--100 MeV, \textbf{bottom-left}: 100 MeV--1 GeV, \textbf{bottom-right}: 1 GeV--10 GeV. Colors refer to different radiator materials. Relative statistical uncertainties are below 0.5\% except for the last energy bin, where they are below 10\%.}
\label{fig:opt_spectra_4pi}
\end{figure*}
\begin{figure}[b]
\centering
\includegraphics[width=.67\columnwidth]{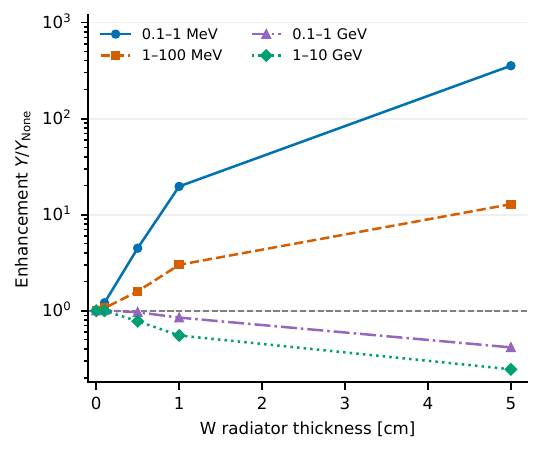}
\caption{Dependence of neutron production enhancement factor on tungsten radiator thickness for four kinetic-energy intervals. Markers denote the simulated thicknesses, $0.1$, $0.5$, $1$ and $5$~cm; the point at zero thickness is the no-radiator reference. The horizontal dashed line marks EF~$=1$.}
\label{fig:tungsten_thickness}
\end{figure}

To resolve the energy dependence of this enhancement, we integrated the neutron production spectra over four kinetic-energy intervals. \Cref{tab:neutron_yields_2} and \Cref{fig:material_enhancement} show the yields and enhancement factors for the four 5~cm radiators. Relative statistical uncertainties on the yields are below 0.5\% in the first three intervals and below 10\% in the 1--10~GeV interval. The enhancement is concentrated below 100~MeV: EF ranges from 170.3 to 548.5 in the 100~keV--1~MeV interval and from 12.9 to 28.8 in the 1--100~MeV interval. At higher energies, all four materials suppress neutron production, with EF~$\approx 0.4$--$0.5$ between 100~MeV and 1~GeV and yields reduced to roughly one quarter to one third of the baseline between 1 and 10~GeV. The corresponding spectral shapes are shown over the full energy range in \Cref{fig:opt_spectra} and separately for each interval in \Cref{fig:opt_spectra_4pi}.

This energy dependence reflects the dual role of the upstream radiator: photonuclear cascades increase the low- and intermediate-energy neutron yield, while changes to the hadronic shower development inside the BD affect the high-energy component. The increase in the total yield reported in \Cref{tab:neutron_yields} is therefore dominated by neutrons below 100~MeV.

Although uranium gives the highest overall enhancement, tungsten offers a more practical alternative because of its lower activation and associated radiation-safety concerns. We therefore extended the tungsten study to thinner radiators, with $\Delta z = 0.1$, $0.5$ and $1$~cm. \Cref{tab:neutron_yields_thickness} reports their yields and enhancement factors alongside the 5~cm result, and \Cref{fig:tungsten_thickness} shows the dependence on thickness. Relative statistical uncertainties are below 0.5\% in the first three energy intervals and below 9\% in the 1--10~GeV interval.

Increasing the tungsten thickness strengthens the low-energy enhancement at the expense of the high-energy component. In the 100~keV--1~MeV interval, EF rises steeply and nonlinearly from approximately 1.2 at 0.1~cm to 19.7 at 1~cm and 355.7 at 5~cm. The 5~cm radiator gives the largest yield in both intervals below 100~MeV among the thicknesses studied, but also the strongest suppression above 100~MeV. Thinner radiators preserve more of the high-energy yield, although at 1~cm the 1--10~GeV yield is already reduced to about 55\% of the baseline.

The preferred thickness thus depends on the energy interval of interest, with radiators of $\Delta z \lesssim 1$~cm offering a compromise between low-energy gain and high-energy attenuation.

\begin{figure*}[t]
    \centering
    \begin{minipage}[t]{0.49\linewidth}
        \centering
        \includegraphics[width=\linewidth]{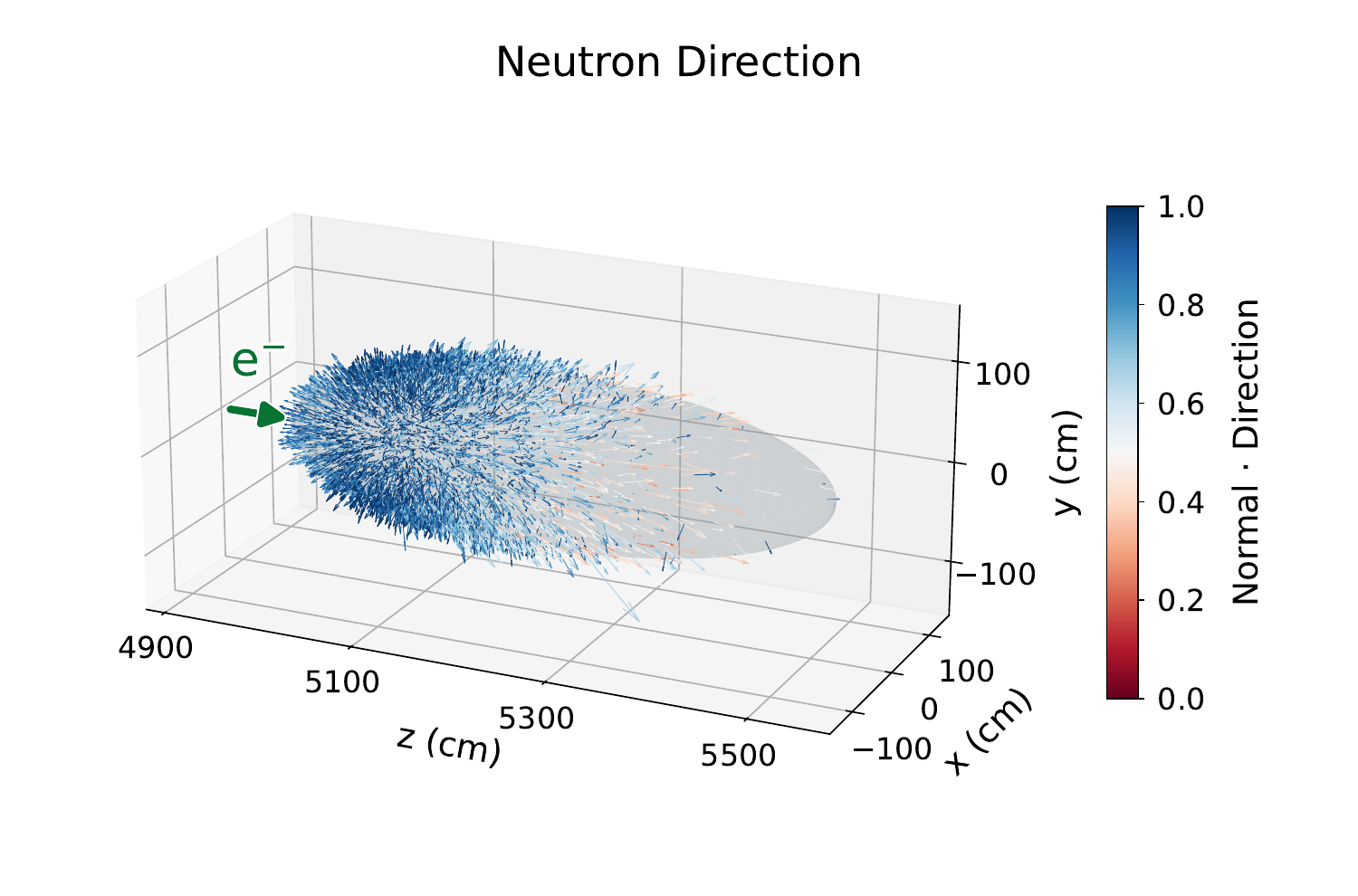}
    \end{minipage}
    \hfill
    \begin{minipage}[t]{0.49\linewidth}
        \centering
        \includegraphics[width=\linewidth]{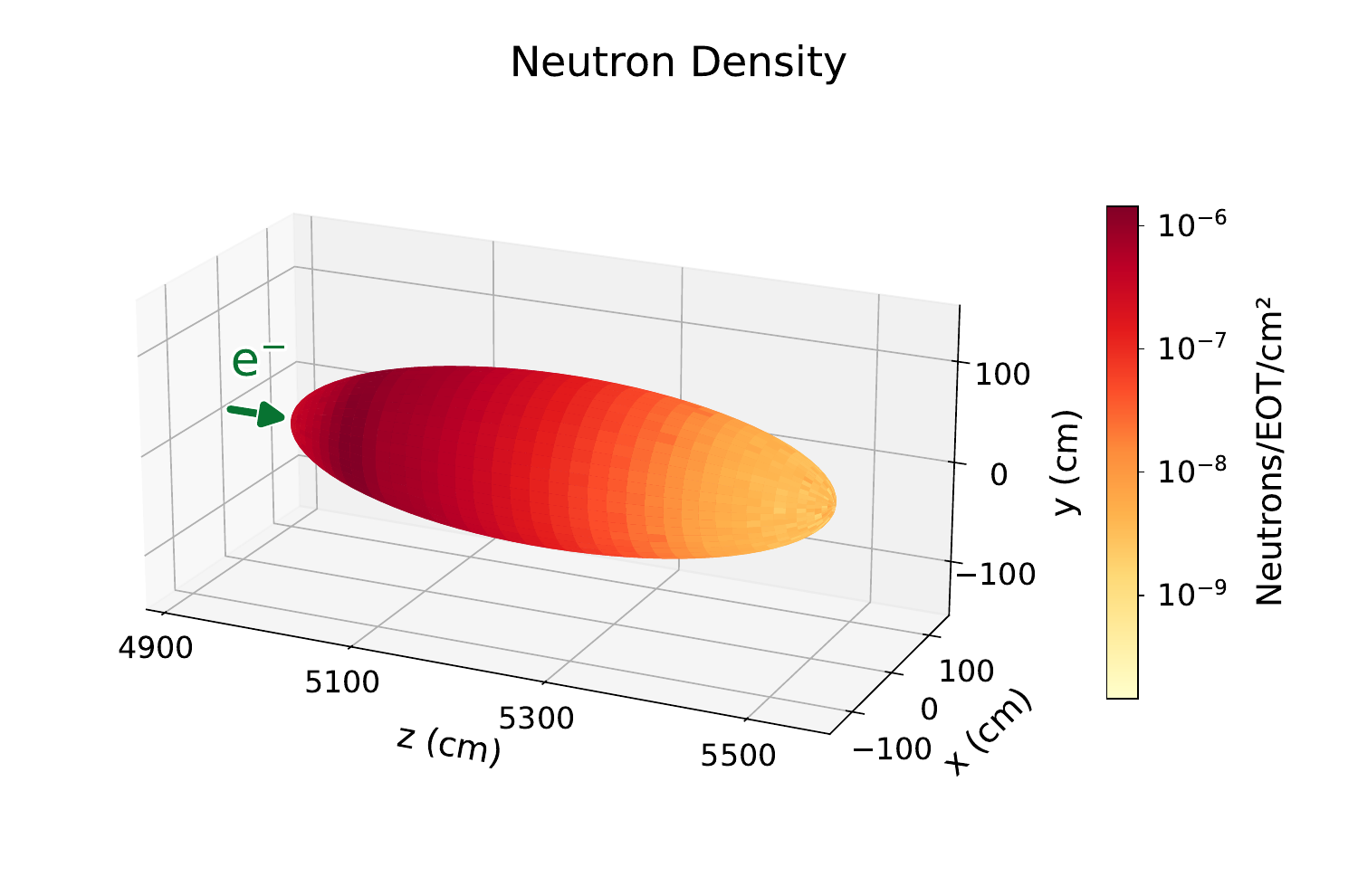}
    \end{minipage}
    \caption{\textbf{Left:} Three-dimensional quiver plot of $N=7000$ neutron crossings sampled uniformly from stored records on the sampling surface (translucent grey ellipsoid) around the BD. Each arrow points along the momentum direction; its length is proportional to $|\vec{P}|$ and its colour encodes the dot product of the outward unit surface normal with the unit neutron direction (NDD), from red (tangential, $\text{NDD}\approx 0$) to blue (along the outward normal, $\text{NDD}\approx 1$). \textbf{Right:} Weighted outgoing neutron crossing density per EOT and per cm$^2$ on the sampling surface, integrated over neutron energy. The green $e^-$ arrow indicates the incident beam direction in both panels.}
    \label{fig:quiver3d_neutrons}\label{fig:quiver3d_density}
\end{figure*}

To characterise the angular distribution of the produced neutrons, we found the distribution in the azimuthal angle $\varphi$ around the beam axis to be approximately uniform in the presence of the radiator, while without it a modulation is observed, reflecting the geometry of the aluminium disks inside the BD, where material is concentrated in the horizontal plane. Since this azimuthal structure is sensitive to the details of the BD geometry as implemented in the simulation, which constitutes an approximation of the real BD, we do not pursue a quantitative characterisation of the $\varphi$ distribution further.

We examined where neutrons cross the ellipsoidal sampling surface. The coordinate $\theta_{\rm ellipse}$ labels crossing positions from downstream of the BD ($0^\circ$) to upstream ($180^\circ$). \Cref{fig:quiver3d_neutrons} illustrates these crossings using a no-radiator sample with $10^6$ EOT.

In the left panel, neutron trajectories are represented by arrows at their crossing positions (a small sample is represented instead of the full statistics). Each arrow points along the neutron momentum, with its length proportional to the momentum magnitude. Their colours encode the cosine of the angle between the neutron momentum and the local outward normal. Values close to unity indicate motion perpendicular to the surface and directed outwards, while values close to zero indicate nearly tangential motion.

The right panel provides the complementary quantitative view: a surface-density map using all selected crossings, weighted and normalised per EOT and per unit surface area.

\begin{figure*}[!htbp]
    \centering
    \includegraphics[width=0.49\linewidth]{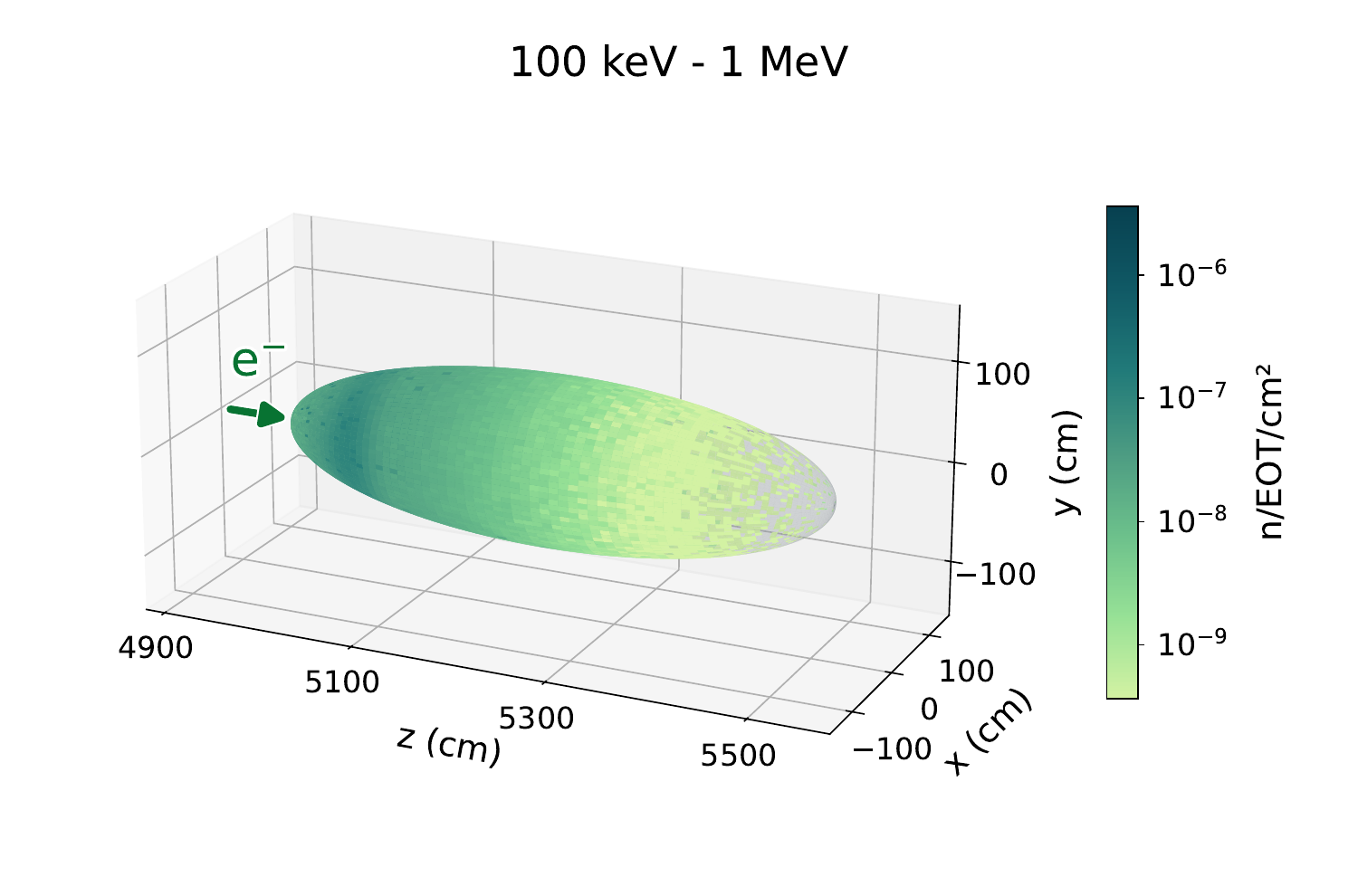}\hfill
    \includegraphics[width=0.49\linewidth]{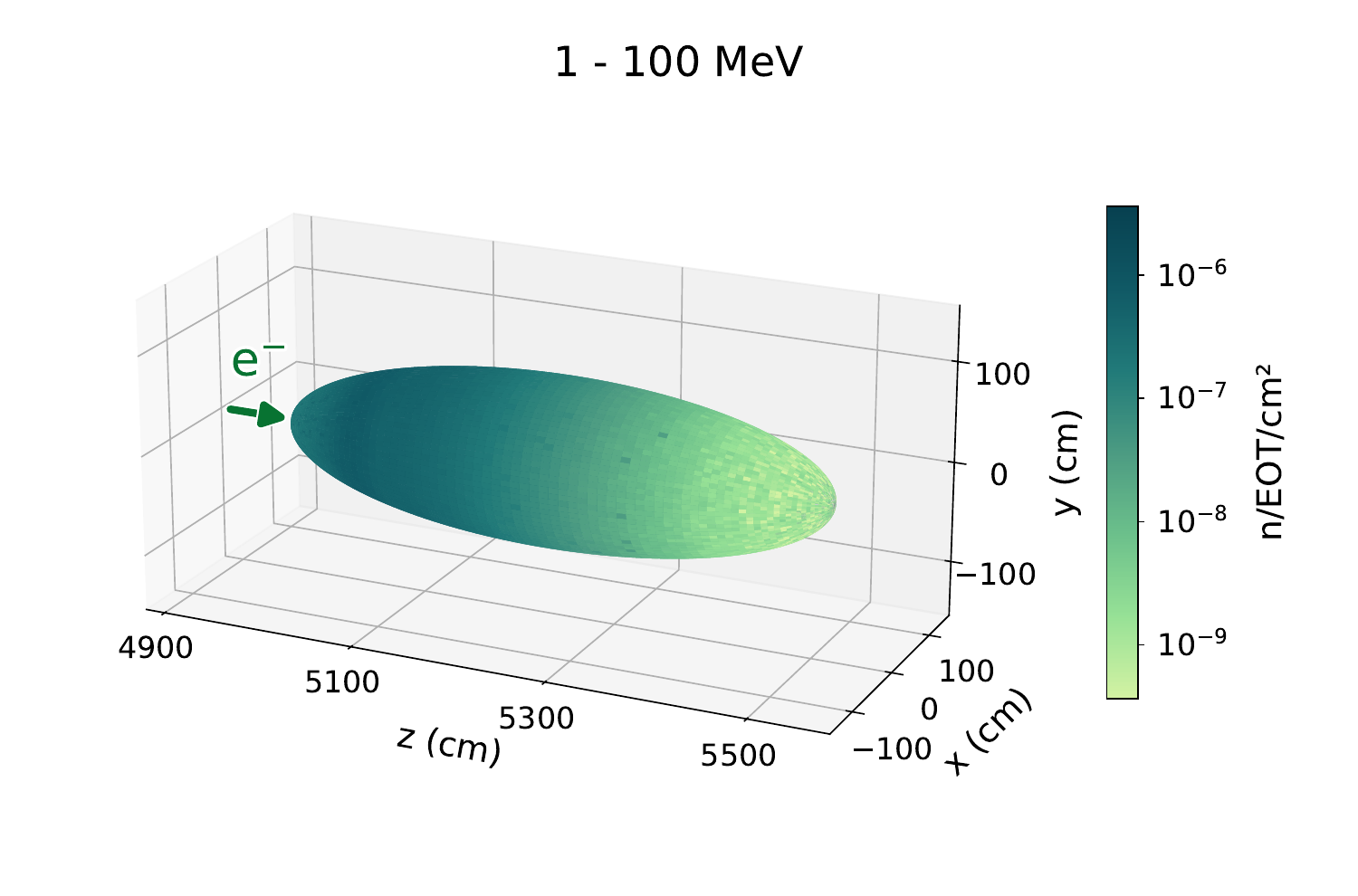}\\
    \includegraphics[width=0.49\linewidth]{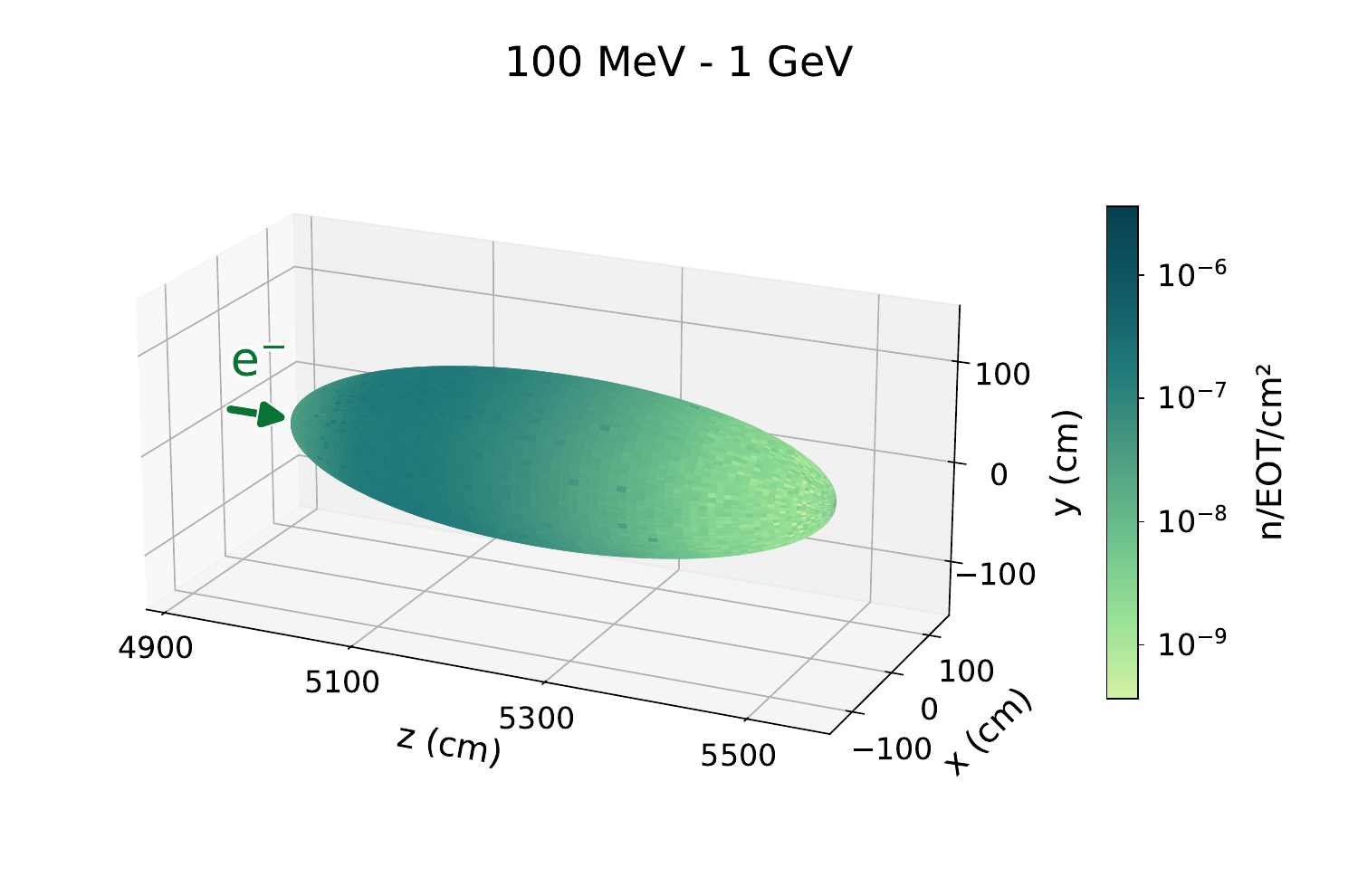}\hfill
    \includegraphics[width=0.49\linewidth]{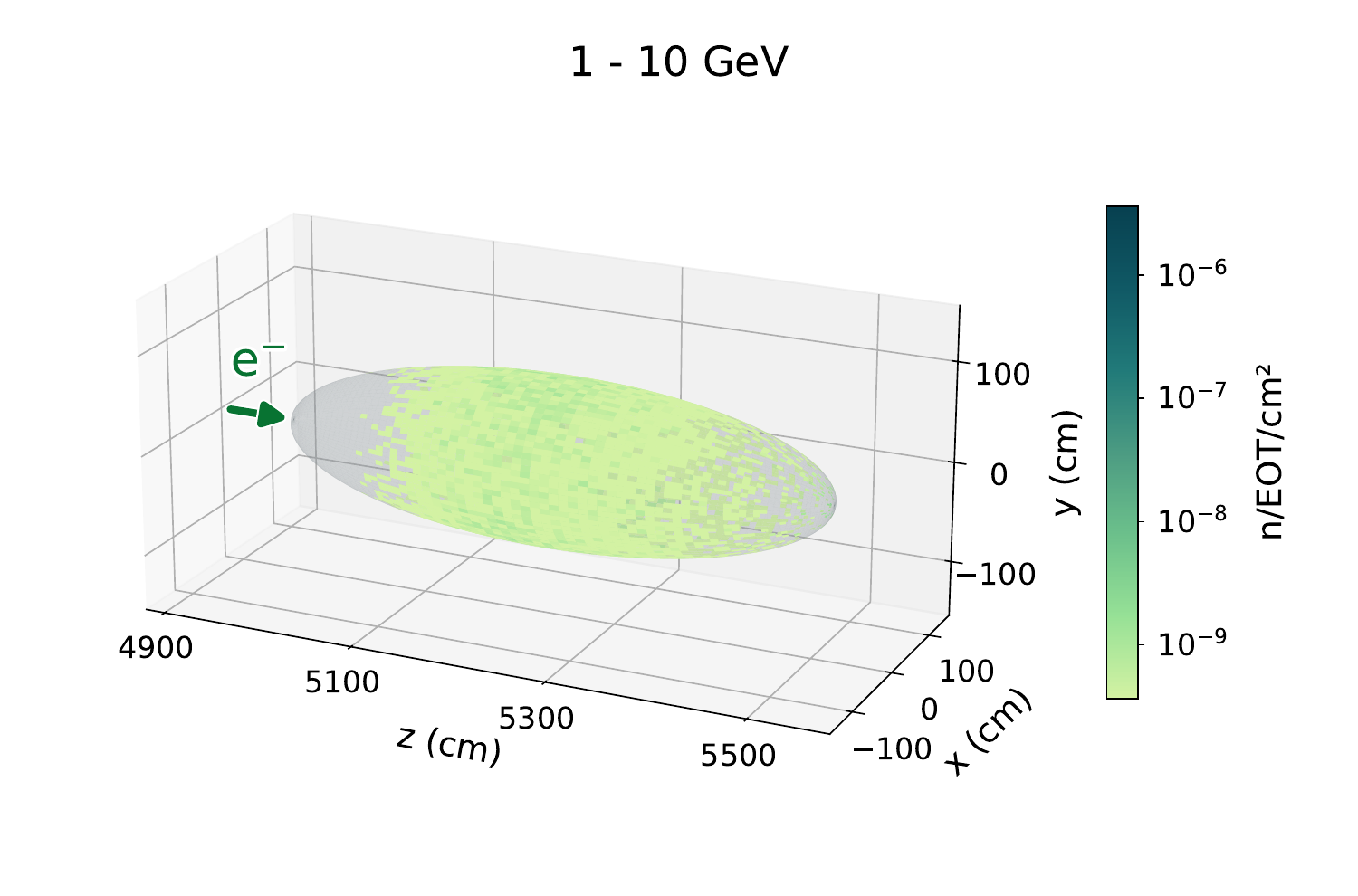}
    \caption{Neutron surface density in n/EOT/cm$^2$ for four kinetic-energy bands in the no-radiator sample ($10^6$ EOT): top-left $100~\text{keV}\leq E_{\rm kin}<1~\text{MeV}$, top-right $1\leq E_{\rm kin}<100$~MeV, bottom-left $100~\text{MeV}\leq E_{\rm kin}<1~\text{GeV}$, bottom-right $1\leq E_{\rm kin}<10$~GeV. The green $e^-$ arrow indicates the incident beam direction.}
    \label{fig:neutron_density_bands}
\end{figure*}

To examine the energy dependence, \Cref{fig:neutron_density_bands} resolves the surface density into four kinetic-energy bands: 100~keV--1~MeV, 1--100~MeV, 100~MeV--1~GeV and 1--10~GeV.
In the three bands below 1~GeV,  neutrons populate both upstream and downstream surface regions, with larger densities on the upstream side. The 1--10~GeV contribution is much weaker on this scale and less dominated by upstream crossings.

\section{Neutron Propagation} \label{prop}
Before reaching locations of interest outside the shielding, neutrons traverse several meters of concrete, soil, and air. 
Because of the strong attenuation associated with thick shielding materials, the neutron fluence decreases by many orders of magnitude as a function of distance, energy, and direction. 
The goal of this section is to quantify such attenuation for the configuration introduced in \Cref{simulations} and to identify positions where a secondary neutron flux may be most effectively exploited. The three geometries considered in this section are shown in \Cref{fig:geom}.

In the following, we first discuss neutron transport for the baseline configuration (A), which corresponds to the current as-built Hall A BD geometry including the two BDX-MINI wells. 
This case provides the most realistic description of the presently available infrastructure. 
We then compare these results with those obtained for Configurations~B and~C, focusing in particular on the flux at the entrance of the BDX facility and on the enhancement achievable through the high-$Z$ pre-target and horizontal extraction pipe of Configuration~C.

The propagation histograms are shown semi-transparently, with a Gaussian-smoothed curve overlaid to convey the underlying spectral trend. 
Far-field yields and derived fluxes with large statistical uncertainties are reported to provide an indicative scale of the flux. Achieving uncertainties comparable to those of the production results would require many more primary electron histories and a computational cost impractical for this study, despite the variance-reduction techniques used.

\subsection{Neutron propagation in Configuration A}\label{subsec:prop_a}

Configuration~A represents the current as-built Hall~A layout, in which no dedicated neutron extraction channel exists. Neutrons must therefore traverse the full thickness of the concrete vault and soil before reaching any accessible location, resulting in severe flux attenuation. This section characterises the neutron field at the positions accessible in the current geometry.

\begin{figure}[!htbp]
\centering
\includegraphics[width=\columnwidth]{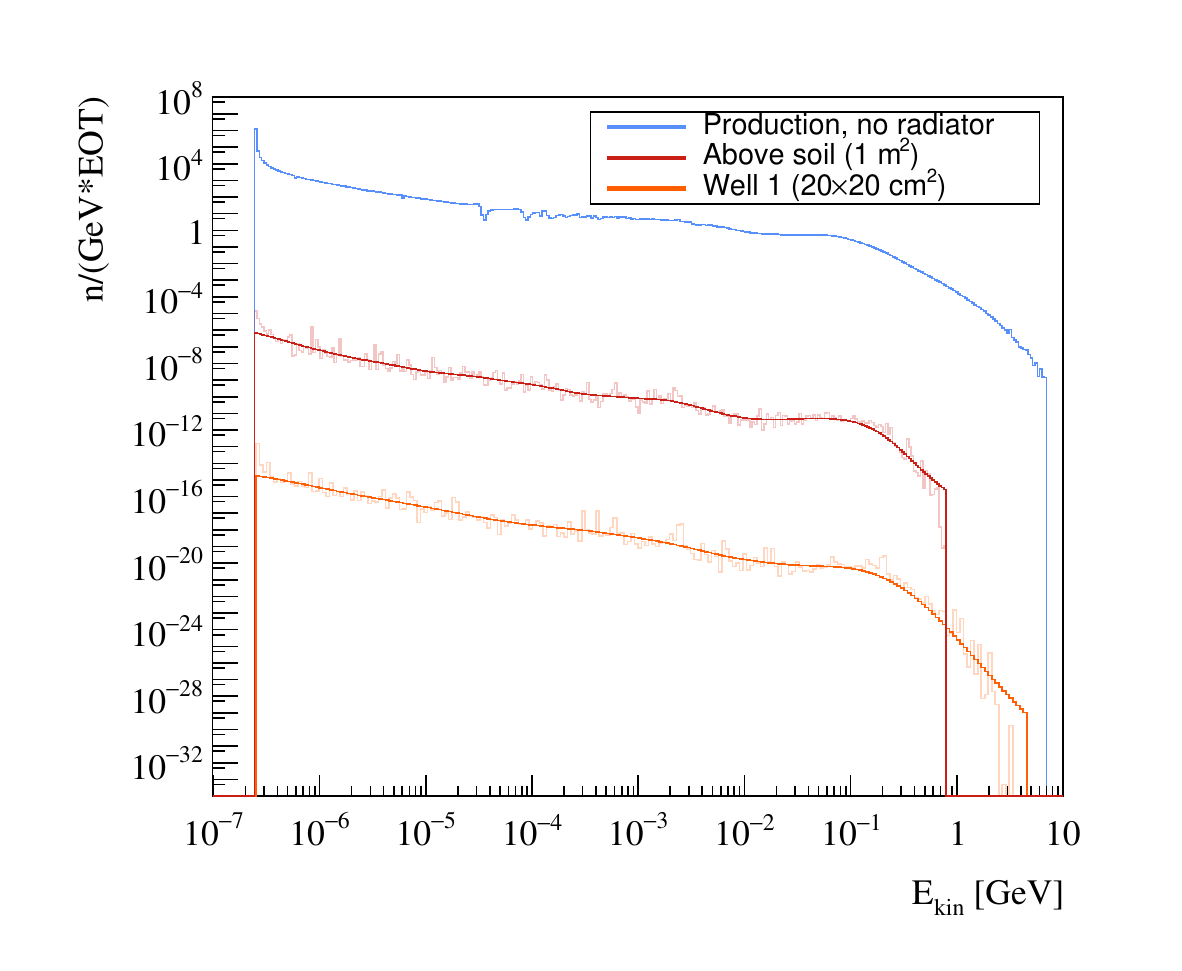}
\caption{Neutron energy spectra for Configuration~A above the soil overburden, integrated over a 1~m$^2$ surface, and at Pipe~1 (BDX-MINI Well~1), integrated over a 20$\times$20~cm$^2$ surface centred on the beam axis, compared with the no-radiator production reference integrated over the ellipsoidal $4\pi$ sampling surface shown in \Cref{fig:bd_geom}. The propagation sampling locations are shown in \Cref{fig:geom}. Pipe~2 is omitted due to insufficient statistics. Semi-transparent propagation histograms are overlaid with Gaussian-smoothed curves in logarithmic energy. These curves are visual guides, not fits.}
\label{fig:propA_spectra_placeholder}
\end{figure}

\Cref{fig:propA_spectra_placeholder} shows the neutron energy spectrum at production (integrated over the full $4\pi$ surface), above the soil overburden (1~m$^2$), and at Pipe~1, one of the two vertical wells installed for the BDX-MINI experiment at $\sim$20~m downstream along the beam axis. Although the neutron flux was sampled over the whole well, the spectrum shown for Pipe~1 is restricted to a $20\times20$~cm$^2$ region centred on the beam axis, i.e.\ the location of maximum expected flux, with the size chosen to match the largest detector that could physically fit inside the well. The spectrum for Pipe~2 is not shown as the extreme neutron attenuation through several metres of concrete and soil reduces the flux to a level where the available statistics are insufficient to obtain a reliable energy-resolved measurement on the $20\times20$~cm$^2$ sampling surface.

Notably, the spectrum sampled above the soil overburden exhibits a sharp drop at energies above approximately 1~GeV. As previously discussed in \Cref{simulations}, this behaviour is a direct consequence of the neutron production kinematics: low-energy neutrons are emitted almost isotropically through nuclear evaporation and multiple scattering, while neutrons above $\sim 1$~GeV arise mainly from deep-inelastic photonuclear reactions and spallation, which strongly boost the secondaries forward along the beam axis. Since the soil overburden lies transverse to the beam axis, at large polar angles, the production of such forward-boosted neutrons is kinematically suppressed in that direction, explaining their absence from the overhead flux.

The total neutron yield integrated over the full energy range above the overburden over a 1 m$^2$ surface is $\sim 2.1 \times 10^{-12}$ n/EOT; while the total neutron yield integrated over the full energy range on the 20$\times$20~cm$^2$ aperture at Pipe~1 is $\sim 1.1 \times 10^{-20}$~n/EOT. The values broken down by energy bin are reported in \Cref{tab:figure4_yields}. Relative statistical uncertainties are 7--12\% for the populated above-soil bins and 7--35\% at Well~1.

\begin{table*}[!htbp]
\centering
\small
\renewcommand{\arraystretch}{1.15}
\setlength{\tabcolsep}{4pt}
\begin{tabular}{@{}l*{4}{c}@{}}
\toprule
\textbf{Location} & \multicolumn{1}{c}{\textbf{100~keV--1~MeV}} & \multicolumn{1}{c}{\textbf{1--100~MeV}} & \multicolumn{1}{c}{\textbf{100~MeV--1~GeV}} & \multicolumn{1}{c}{\textbf{1--10~GeV}} \\
\midrule
Production (no rad.) & $5.7 \times 10^{-3}$ & $6.3 \times 10^{-2}$ & $1.8 \times 10^{-2}$ & $5.6 \times 10^{-5}$ \\
Above soil (100$\times$100~cm$^2$) & $1.7 \times 10^{-13}$ & $7.8 \times 10^{-13}$ & $3.8 \times 10^{-13}$ & $0.0$ \\
BDX-Mini Well~1 (20$\times$20~cm$^2$) & $1.5 \times 10^{-21}$ & $2.3 \times 10^{-21}$ & $1.6 \times 10^{-21}$ & $5.1 \times 10^{-25}$ \\
\bottomrule
\end{tabular}
\caption{Neutron yields per EOT for different detector locations and energy bins in geometry configuration A. The first row shows the production yield at the BD sampling surface enclosure (reference). The second row shows the yield above the soil surface (100$\times$100~cm$^2$ area). The third row shows the yield at BDX-Mini Well~1 (20$\times$20~cm$^2$ area). Relative statistical uncertainties, in order of increasing energy, are approximately 12\%, 7\%, and 9\% in the first three above-soil bins, and 35\%, 21\%, 7\%, and 21\% at Well~1. The last energy bin for the above-soil detector yielded no counts in the available simulation sample, so no relative statistical uncertainty is assigned to that entry. The large uncertainties reflect the limited statistics at the far flux-detectors. We do not report neutrons below 100 keV.
}
\label{tab:figure4_yields}
\end{table*}

Without a dedicated extraction channel, the neutron flux decreases by many orders of magnitude relative to the production value. To assess the practical impact, one can consider the expected operating conditions of the upcoming MOLLER experiment in Hall A at JLab~\cite{moller1,moller2}, which will use the same 11~GeV electron beam at a current of $I = 65~\mu$A, corresponding to a primary beam rate of $\sim 4.06\times10^{14}$~EOT/s, or approximately $10^{22}$~EOT/year. At yearly rate, the expected yield at Pipe~1 is only about $1.1\times10^{2}$ neutrons per year, even assuming ideal detection efficiency and negligible background. This low flux motivates a dedicated neutron extraction geometry, as pursued in Configuration~C through the horizontal extraction pipe.

In terms of instantaneous counting rates, at this same beam current the expected neutron flux above the soil overburden would be $N_{\rm Soil} \approx 8.7\times10^{2}$~n/s over the whole spectra, while the rate on the $20\times20$~cm$^2$ surface placed inside BDX-MINI Well~1, centred on the beam axis, would amount to only $N_{\text{Well-1}} \approx 4.5\times10^{-6}$~n/s.

\subsection{Propagation in Configurations B and C}

Configurations~B and~C share the same downstream BDX facility geometry. Configuration~B has no dedicated extraction channel, whereas Configuration~C includes a horizontal air-filled pipe, 50.8~cm (20~in) in diameter, connecting the BD concrete vault to the facility. Each configuration is simulated with and without a 5~cm tungsten radiator upstream of the BD. This comparison separates the effect of the extraction pipe on neutron transport from that of the radiator on the neutron source.

\begin{figure*}[t]
\centering
    \centering
    \begin{minipage}[t]{0.49\linewidth}
        \centering
        \includegraphics[width=\linewidth]{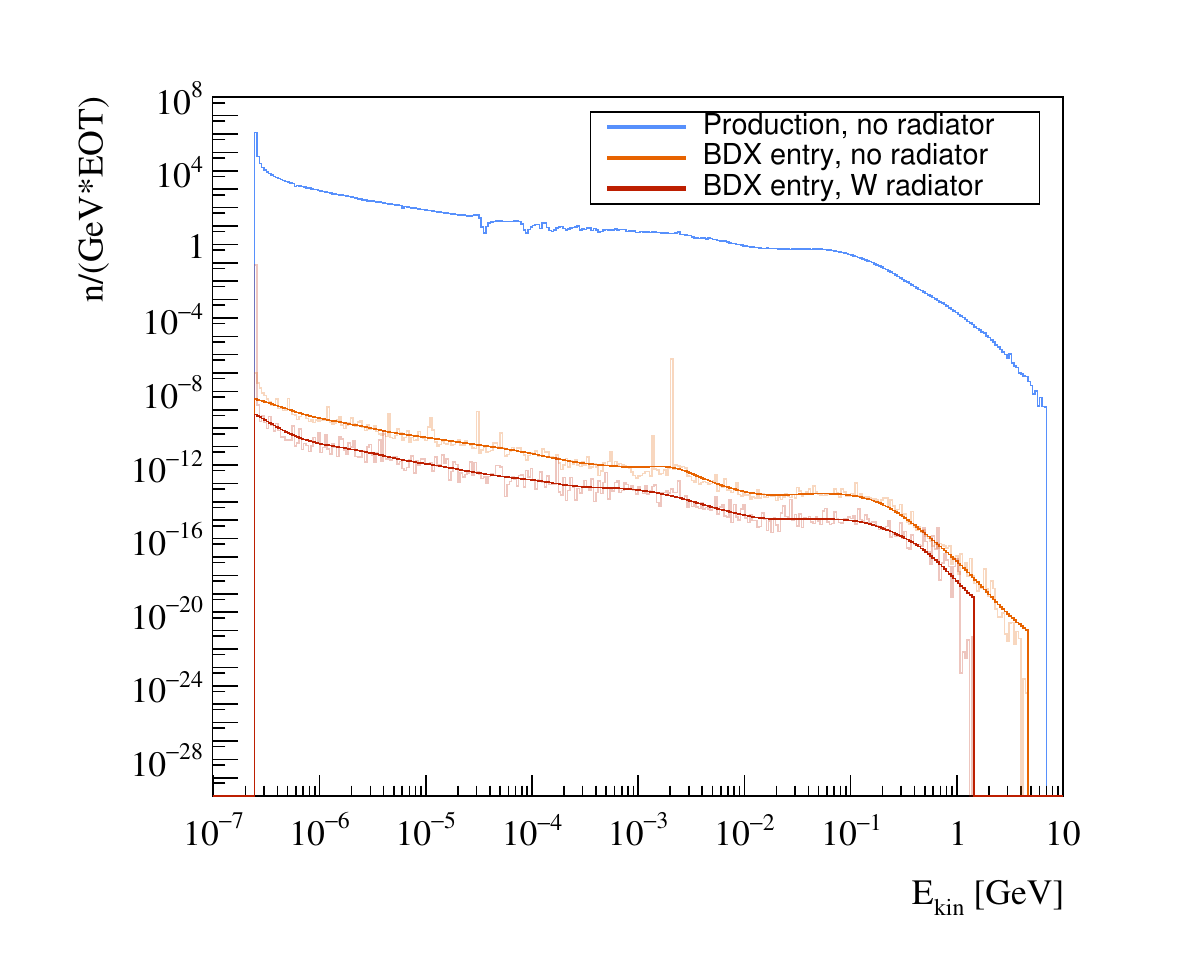}
    \end{minipage}
    \hfill
    \begin{minipage}[t]{0.49\linewidth}
        \centering
        \includegraphics[width=\linewidth]{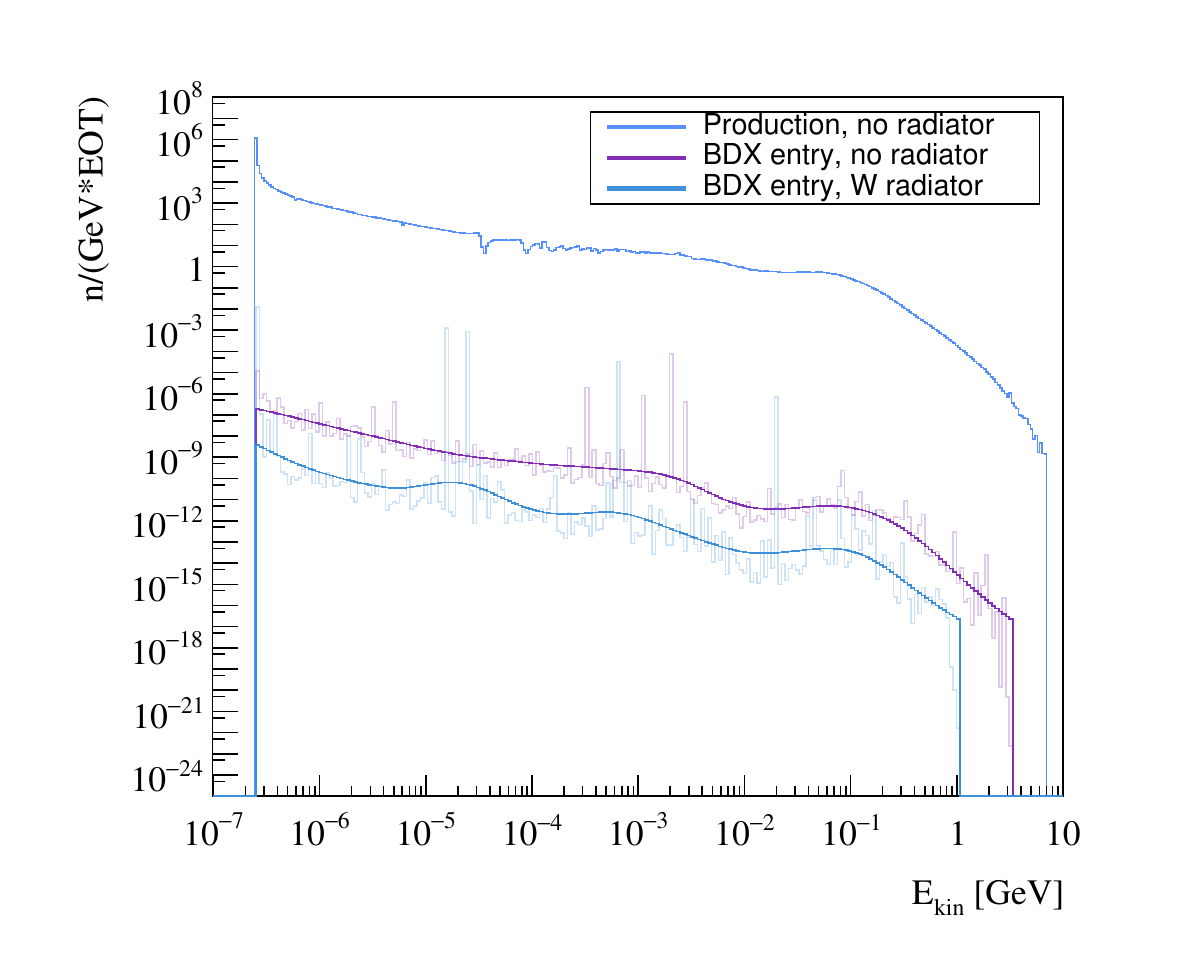}
    \end{minipage}
\caption{Neutron energy spectra for configurations~B and~C at the BDX facility entrance, integrated over a 1~m$^2$ surface centred on the beam axis, compared with the no-radiator production reference integrated over the ellipsoidal $4\pi$ sampling surface shown in \Cref{fig:bd_geom}. The propagation sampling locations are shown in \Cref{fig:geom}. \textbf{Left:} Configuration~B (without the extraction pipe), with and without the 5~cm tungsten radiator. \textbf{Right:} Configuration~C (with the extraction pipe), with and without the same radiator. Semi-transparent propagation histograms are overlaid with Gaussian-smoothed curves in logarithmic energy. These curves are visual guides, not fits.}
\label{fig:propBC_compare_placeholder}
\end{figure*}

In all four cases, neutrons are sampled over a 1~m$^2$ surface centred on the beam axis at the BDX facility entrance. \Cref{fig:propBC_compare_placeholder} compares these spectra with the no-radiator production spectrum integrated over the $4\pi$ sampling surface surrounding the BD. The production spectrum provides a reference for the overall yield reduction between the source and the facility entrance; the two sampling surfaces have different geometrical acceptances.

\begin{table*}[t]
\centering
\small
\renewcommand{\arraystretch}{1.15}
\setlength{\tabcolsep}{4pt}
\begin{tabular}{@{}l*{4}{c}@{}}
\toprule
\textbf{Configuration} & \multicolumn{1}{c}{\textbf{100~keV--1~MeV}} & \multicolumn{1}{c}{\textbf{1--100~MeV}} & \multicolumn{1}{c}{\textbf{100~MeV--1~GeV}} & \multicolumn{1}{c}{\textbf{1--10~GeV}} \\
\midrule
Production & $5.7 \times 10^{-3}$ & $6.3 \times 10^{-2}$ & $1.8 \times 10^{-2}$ & $5.6 \times 10^{-5}$ \\
Config~B (No rad.) & $1.2 \times 10^{-15}$ & $6.0 \times 10^{-11}$ & $3.3 \times 10^{-15}$ & $2.7 \times 10^{-18}$ \\
Config~B (W rad.) & $9.6 \times 10^{-17}$ & $2.5 \times 10^{-16}$ & $1.9 \times 10^{-16}$ & $3.1 \times 10^{-21}$ \\
\bottomrule
\end{tabular}
\caption{Neutron yields at the BDX facility entrance (1~m$^2$) per EOT for different energy bins in geometry configuration B. The first row shows the production yield at the BD sampling surface enclosure (reference). Relative statistical uncertainties, in order of increasing energy, are approximately 11\%, 71\%, 8\%, and 24\% without the radiator, and 25\%, 10\%, 10\%, and 28\% with the W radiator. The large uncertainties reflect the limited statistics at the far flux-detector. We do not report neutrons below 100 keV.}
\label{tab:spectrum_yields_config_B}
\begin{tabular}{@{}l*{4}{c}@{}}
\toprule
\textbf{Configuration} & \multicolumn{1}{c}{\textbf{100~keV--1~MeV}} & \multicolumn{1}{c}{\textbf{1--100~MeV}} & \multicolumn{1}{c}{\textbf{100~MeV--1~GeV}} & \multicolumn{1}{c}{\textbf{1--10~GeV}} \\
\midrule
Production & $5.7 \times 10^{-3}$ & $6.3 \times 10^{-2}$ & $1.8 \times 10^{-2}$ & $5.6 \times 10^{-5}$ \\
Config~C (No rad.) & $5.1 \times 10^{-11}$ & $1.2 \times 10^{-8}$ & $1.1 \times 10^{-12}$ & $6.2 \times 10^{-15}$ \\
Config~C (W rad.) & $1.6 \times 10^{-9}$ & $1.1 \times 10^{-9}$ & $6.8 \times 10^{-14}$ & $2.2 \times 10^{-17}$ \\
\bottomrule
\end{tabular}
\caption{Neutron yields at the BDX facility entrance (1~m$^2$) per EOT for different energy bins in geometry configuration C. The first row shows the production yield at the BD sampling surface enclosure (reference). Relative statistical uncertainties, in order of increasing energy, are approximately 100\% for the first two energy bins and 17\% and 26\% for the other two without the radiator, and 100\%, 71\%, 32\%, and 50\% with the W radiator. The large uncertainties reflect the limited statistics at the far flux-detector. We do not report neutrons below 100 keV.}
\label{tab:spectrum_yields_config_C}
\end{table*}

The extraction pipe increases the nominal yield at the facility entrance in all four reported energy intervals, both with and without the radiator. It provides an air-filled path through the shielding for neutrons travelling towards the facility, reducing the amount of material they must traverse. This path is particularly relevant to the forward-directed component of the source spectrum.

The effect of the tungsten radiator depends on both the transport geometry and the neutron energy. In Configuration~B, adding the radiator reduces the nominal yield in every reported interval from 100~keV to 10~GeV. Thus, the increased neutron production discussed above does not translate into a larger yield at the facility entrance in these intervals. The downstream spectra reflect the combined effects of neutron production and transport through the shielding.

In Configuration~C, adding the radiator increases the nominal yield in the 100~keV--1~MeV interval and reduces it in the three higher-energy intervals. The radiator therefore favours the sub-MeV component in the nominal spectra, while the extraction pipe improves neutron delivery with either source configuration. The magnitude of these changes remains uncertain, particularly in the two lowest-energy intervals of Configuration~C, where relative statistical uncertainties range from approximately 71\% to 100\%.

The yields per EOT in the four energy intervals are listed in \Cref{tab:spectrum_yields_config_B} and~\Cref{tab:spectrum_yields_config_C}, with relative statistical uncertainties in the captions. These tables cover 100~keV--10~GeV; the full-spectrum totals discussed below also include neutrons below 100~keV.

The full-spectrum yields are obtained by integrating the dedicated spectra over 0.1~eV--10~GeV on the same 1~m$^2$ entrance surface. They are therefore not the sums of the four tabulated energy intervals. For the 11~GeV beam and $I=65~\mu$A operating conditions introduced in \Cref{subsec:prop_a}, the corresponding neutron rates are calculated as $N_{\rm BDX}=Y(I/e)$, with $I/e \approx 4.06\times10^{14}$~EOT/s:

\begin{table*}[t]
\centering
\small
\renewcommand{\arraystretch}{1.15}
\setlength{\tabcolsep}{4pt}
\begin{tabular}{@{}l*{4}{c}@{}}
\toprule
\textbf{Configuration} & \multicolumn{1}{c}{\textbf{100~keV--1~MeV}} & \multicolumn{1}{c}{\textbf{1--100~MeV}} & \multicolumn{1}{c}{\textbf{100~MeV--1~GeV}} & \multicolumn{1}{c}{\textbf{1--10~GeV}} \\
\midrule
Production & $2.3 \times 10^{12}$ & $2.5 \times 10^{13}$ & $7.3 \times 10^{12}$ & $2.3 \times 10^{10}$ \\
Config~B (No rad.) & $4.9 \times 10^{-1}$ & $2.4 \times 10^{4}$ & $1.4$ & $1.1 \times 10^{-3}$ \\
Config~B (W rad.) & $3.9 \times 10^{-2}$ & $1.0 \times 10^{-1}$ & $7.8 \times 10^{-2}$ & $1.3 \times 10^{-6}$ \\
\bottomrule
\end{tabular}
\caption{Neutron rates (n/s) at the BDX facility entrance (1~m$^2$) for different energy bins in geometry configuration B. The first row shows the production rate at the sampling surface surrounding the BD (reference). The rates are computed assuming $E = 11~\text{GeV}$ and $I = 65~\mu$A ($I/e \approx 4.06\times10^{14}$~e$^{-}$/s). The relative statistical uncertainties are the same as those of the corresponding yields, listed in the caption of \Cref{tab:spectrum_yields_config_B}. The large uncertainties reflect the limited statistics at the far flux-detector. We do not report neutrons below 100 keV.}
\label{tab:rates_config_B}
\begin{tabular}{@{}l*{4}{c}@{}}
\toprule
\textbf{Configuration} & \multicolumn{1}{c}{\textbf{100~keV--1~MeV}} & \multicolumn{1}{c}{\textbf{1--100~MeV}} & \multicolumn{1}{c}{\textbf{100~MeV--1~GeV}} & \multicolumn{1}{c}{\textbf{1--10~GeV}} \\
\midrule
Production & $2.3 \times 10^{12}$ & $2.5 \times 10^{13}$ & $7.3 \times 10^{12}$ & $2.3 \times 10^{10}$ \\
Config~C (No rad.) & $2.1 \times 10^{4}$ & $4.9 \times 10^{6}$ & $4.3 \times 10^{2}$ & $2.5$ \\
Config~C (W rad.) & $6.6 \times 10^{5}$ & $4.4 \times 10^{5}$ & $2.7 \times 10^{1}$ & $8.9 \times 10^{-3}$ \\
\bottomrule
\end{tabular}
\caption{Neutron rates (n/s) at the BDX facility entrance (1~m$^2$) for different energy bins in geometry configuration C. The first row shows the production rate at the sampling surface surrounding the BD (reference). The rates are computed assuming $E = 11~\text{GeV}$ and $I = 65~\mu$A ($I/e \approx 4.06\times10^{14}$~e$^{-}$/s). The relative statistical uncertainties are the same as those of the corresponding yields, listed in the caption of \Cref{tab:spectrum_yields_config_C}. The large uncertainties reflect the limited statistics at the far flux-detector. We do not report neutrons below 100 keV.}
\label{tab:rates_config_C}
\end{table*}
\begin{itemize}
    \item Configuration~B, without radiator: $Y \approx 6.0\times10^{-11}$~n/EOT, corresponding to $N_{\rm BDX} \approx 2.4\times10^{4}$~n/s.
    \item Configuration~B, with radiator: $Y \approx 9.4\times10^{-10}$~n/EOT, corresponding to $N_{\rm BDX} \approx 3.8\times10^{5}$~n/s.
    \item Configuration~C, without radiator: $Y \approx 1.2\times10^{-8}$~n/EOT, corresponding to $N_{\rm BDX} \approx 4.9\times10^{6}$~n/s.
    \item Configuration~C, with radiator: $Y \approx 6.1\times10^{-9}$~n/EOT, corresponding to $N_{\rm BDX} \approx 2.5\times10^{6}$~n/s.
\end{itemize}

The distinction between full-spectrum totals and yields above 100~keV is essential for interpreting the radiator's effect. In Configuration~B, the larger nominal full-spectrum yield with the radiator is dominated by neutrons below 100~keV, even though every tabulated yield above that threshold decreases. In Configuration~C, the radiator raises the nominal rate in the 100~keV--1~MeV interval from approximately $2.1\times10^{4}$ to $6.6\times10^{5}$~n/s, while lowering the full-spectrum rate. The preferred source configuration therefore depends on the energy interval required by the application. These comparisons indicate nominal trends rather than precise enhancement factors, given the large statistical uncertainties in several energy intervals. The rates in each interval are listed in \Cref{tab:rates_config_B} and~\Cref{tab:rates_config_C}.

\subsection{Comparison with neutron facilities}\label{subsec:beam_comparison}

The proposed BDX source would complement existing neutron facilities through access to neutrons above 1~GeV during routine electron-beam operation. In the 1--100~MeV interval, published energy-integrated fluxes are approximately $4\times10^{6}$~n$/$cm$^2$s at CHIPIR~\cite{Cazzaniga:2024ChipIr}, $1\times10^{7}$~n/cm$^2$s at TRIUMF TNF~\cite{Monk:2015TNF}, $5\times10^{5}$~n/cm$^2$s at TRIUMF BL1B~\cite{BelangerChampagne:2024PIFNIF}, and $3\times10^{6}$~n/cm$^2$s at CSNS Back-n~\cite{Chen:2019Backn}, under their reported operating conditions. In their respective reference intervals above 100~MeV, CHIPIR and TRIUMF provide approximately $2$--$30\times10^{5}$~n/cm$^2$s. Dedicated sources also offer higher intensities at lower energies, including the HFIR irradiation regions~\cite{Skutnik:2023HFIR}. These selected reference intervals extend up to 800~MeV, the highest endpoint being that of CHIPIR. BDX is not competitive in intensity within these overlapping ranges.

For an 11~GeV electron beam at 65~$\mu$A, Configuration~C without the radiator gives a nominal rate of approximately 2.5~n/s in the 1--10~GeV interval, integrated over the 1~m$^2$ surface at the facility entrance. The opportunity is to deliver this component through the horizontal extraction pipe alongside the primary electron-scattering programme, without dedicated accelerator running.

\section{Conclusions}\label{conclusions}

We investigated the prospects for secondary neutron beams at Jefferson Lab using FLUKA simulations of neutron production by the 11~GeV electron beam in the experimental Hall~A BD. The predicted spectrum extends from thermal energies to several GeVs. Although the fluxes are substantially lower than those of established neutron facilities in the overlapping energy intervals examined, this source offers access to neutrons above 1~GeV during normal operation of the primary electron-scattering experiment.

Exploiting this component requires a cavern downstream of the BD vault and a horizontal extraction pipe through the soil shielding, which otherwise strongly attenuates the downstream neutron flux. With this extraction channel and no upstream radiator, the predicted rate at the facility entrance is approximately 2.5~neutrons/s in the 1--10~GeV interval over an area of 1~m$^2$, for a beam current of 65~$\mu$A.
We also studied configurations with an upstream radiator to adjust the spectrum for specific applications. A 5~cm tungsten radiator increases the 100~keV--1~MeV component, while omitting it preserves the higher-energy yields. Radiator selection is therefore secondary to establishing the extraction channel.

\begin{acknowledgments}
We thank Lorenzo~Zana for his support during the early stages of this work. We thank Andrea~Celentano for providing the initial geometry model and simulation codes that served as a starting point for this study. We are  grateful to Walt~Akers for his assistance in verifying the Hall A BD geometry and for his continued support throughout the project. We also acknowledge the computing resources provided by Thomas Jefferson National Accelerator Facility, which were used to perform all simulations presented in this work. This material is based upon work supported by the U.S. Department of Energy, Office of Science, Office of Nuclear Physics under Contract No.~89243126CSC000213. AF acknowledges financial support under the program ``Progetti di Rilevante Interesse Nazionale'' (PRIN 2022), published on 2.2.2022 by the Italian Ministry of University and Research (MUR), Project Title ``The X(3872) files'' -- CUP J53C24002600006 -- Grant Assignment Decree No. 20429 adopted on 6.11.2024 by the Italian Ministry of University and Research (MUR).
\end{acknowledgments}

\bibliographystyle{bst/apsrev4-2}
\bibliography{biblio}

\end{document}